\documentstyle[preprint,epsfig,aps,amssymb]{revtex}
\begin{document}
\draft
\tightenlines

\def\br{{\bf r}}
\def\bk{{\bf k}}
\def\bz{{\bf z}}
\def\brt{\br,t}
\def\bbrt{(\brt)}
\def\bprt{\bp,\brt}
\def\cphio{\Phi_0}
\def\bna{\bbox{\nabla}}
\def\bp{{\bf p}}
\def\bv{{\bf v}}
\def\tn{\tilde n}
\def\tp{\tilde p}
\def\be{\bbox{\eta}}

\title{Modelling Bose-condensed gases at finite temperatures with
 $N$-body simulations}
\author{B. Jackson and E. Zaremba}
\address{Department of Physics, Queen's University, Kingston, Ontario 
 K7L 3N6, Canada.}
\date{\today}
\maketitle
\begin{abstract}
We consider a model of a dilute Bose-Einstein condensed gas at finite 
temperatures, where the condensate coexists in a trap with a cloud
of thermal excitations. Within the ZGN formalism, the dynamics of the 
condensate is described by a generalized Gross-Pitaevskii equation, 
while the thermal cloud is represented by a semiclassical kinetic 
equation. Our numerical approach 
simulates the kinetic equation using a cloud of representative
test particles, while collisions are treated by means of a Monte Carlo 
sampling technique. A full 3D split-operator Fast Fourier Transform 
method is used to evolve the condensate wavefunction.
We give details regarding the numerical methods
used and discuss simulations carried out to test the accuracy of the
numerics. We use this scheme to simulate the monopole
mode in a spherical trap. The dynamical coupling between the condensate
and thermal cloud is responsible for frequency shifts and damping of
the condensate collective mode. We compare our results to previous
theoretical approaches, not only to confirm the reliability of our
numerical scheme, but also to check the validity of approximations 
which have been used in the past.

\end{abstract}

\section{Introduction}

Bose-Einstein condensation (BEC), whereby bosons form a condensate by
macroscopically occupying the lowest energy state of the system,
is a striking and important consequence
of quantum statistics at low temperatures.  The resultant 
long-range order manifests itself in phenomena such as macroscopic
coherence and superfluidity.
In general the condensate is depleted by correlation
effects and through thermal population of excited states at finite 
temperatures. The former, termed {\it quantum depletion}, is particularly
important for dense fluids such as liquid $^4{\rm He}$, where only around 
10\% of the atoms are condensed
in the low temperature limit. In contrast, the quantum depletion in
trapped, dilute gaseous BECs \cite{anderson95,bradley95,davis95}
is typically less than 1\% 
\cite{dalfovo99,hutchinson97}. The noncondensed fraction is thus
mainly composed of thermal excitations, and almost pure condensates can 
be prepared by evaporative cooling to very low temperatures. Atomic vapors
therefore allow unique opportunities to study the properties of Bose
condensates under a wide range of conditions, from the pure condensate
phase to the noncondensed thermal cloud above the BEC transition. 

The condensate in a dilute Bose gas is well-described by means of
a macroscopic
wavefunction, which in the limit of low temperatures evolves according 
to the Gross-Pitaevskii (GP) equation. Well-known techniques allow both 
numerical and analytical solutions of this equation, and comparisons
with experiment at low temperatures show excellent agreement for both 
static 
and dynamical properties \cite{dalfovo99}. However, generalizations of
the theory to finite temperatures, where thermal excitations coexist 
with the condensate, have proved far more difficult.
To accurately describe the dynamical behavior in this situation
requires a theory which treats both
components in a fully consistent manner. Such theories have recently
been formulated, but the challenge of obtaining explicit solutions has
remained. What has been lacking in particular is a computationally
feasible method for treating the dynamics of the thermal cloud.
It is these computational aspects that concern us most in this paper.


The earliest studies of dynamics at finite temperatures were based on 
the Hartree-Fock-Bogoliubov (HFB) approximation
\cite{hutchinson97,dodd98}. Within this theory, excitations of the
condensate are obtained by solving the HFB equations which are derived 
by linearizing
the GP equation about the equilibrium solution, or equivalently,
from the grand canonical Hamiltonian of the system
\cite{griffin96}. The frequencies of the excitations are identified with
the collective modes of the condensate. This theory, however, is
incomplete. Although the excitations are thermally populated, the
condensate in fact oscillates in the presence of a {\it static} thermal
cloud. This ignores the dynamical response of the thermal cloud to
condensate fluctuations which is responsible for Landau damping and
associated frequency shifts. By the same token,
the theory cannot be used to account for the response of the system to
external perturbations as typically used in experiments to excite the
trapped gas \cite{jin97,stamper-kurn98,marago01,chevy01}.
This problem becomes critical at high temperatures, where 
collective motion of the thermal cloud can exert a major influence on 
the condensate evolution, as reflected in experimental results for the 
mode frequency and damping rate.

Recent important work by Morgan 
{\it et al.}\cite {morgan00,rusch00} and Giorgini 
\cite{giorgini98,giorgini00}
has extended the HFB theory to include collisionless noncondensate
dynamics within second-order perturbation schemes, and derived 
expressions for damping rates and frequency shifts of low-energy modes.
A variant of these approaches is the dielectric response formulation of
Reidl {\it et al.} \cite{reidl00}. One limitation of these theories is
the absence of collisions which require a kinetic theory for their
description. Quantum kinetic equations for BECs 
have been developed by Gardiner and collaborators
\cite{gardiner00}, Stoof \cite{stoof99}
and Walser {\it et al.} \cite{walser99}. 
However, calculations based on these theories are very difficult to
carry out and as a result, they have not yet been used to study
collective excitations.
A somewhat simpler scheme is the one developed by Zaremba, Griffin and 
Nikuni (ZGN) \cite{nikuni99,zaremba99}, which treats the
excitations semiclassically within the Hartree-Fock (HF) and Popov
approximations. One can then identify the excitations with a thermal cloud of
particles, with dynamics governed by a Boltzmann equation for the phase-space 
density. In 
analogy with its classical counterpart, binary collisions between particles 
are described by means of a collision integral; however, an additional 
collision integral arises to account for collisions
with the condensate. The latter leads to an important modification of
the GP equation which must now include a
non-Hermitian source term to account for the transfer of atoms 
into and out of the condensate. This process, taken together 
with mean field coupling between the two components, leads to damping
and frequency shift of the condensate collective modes at finite 
temperature.

The coupled GP and Boltzmann equations are far from trivial to solve, and 
several approximations have been invoked in the literature in order to explore
their properties. When the characteristic collisional time scale, $\tau$, 
satisfies $\omega_0 \tau \ll 1$, where $\omega_0$ is the trap frequency, then 
collisions dominate and the system is said to be in the 
hydrodynamical regime. One can then take moments of the kinetic equation
to derive a set of coupled hydrodynamic equations for the 
noncondensate which, together with similar equations for the condensate,
can be solved under certain conditions 
\cite{zaremba99,nikuni01a}. In the opposite collisionless regime, 
$\omega_0 \tau \gg 1$, 
Stoof and co-workers
\cite{bijlsma99,alkhawaja00} have used a joint variational and moment 
scheme to model the condensate and noncondensate, respectively, while 
Nikuni \cite{nikuni01b} recently applied a moment method to study 
the scissors mode \cite{guery-odelin99,marago00,marago01}. 
Although these moment methods provide some insight into the coupled 
dynamics of the two components, they constitute a truncated description
which precludes coupling to internal degrees of freedom of the
gas. Thus, Landau damping is neglected.
In order to avoid this limitation, and to facilitate direct 
comparisons with experiment, one must resort to the full kinetic theory.
It is therefore desirable to directly simulate the ZGN equations without
making approximations beyond those used to derive the equations
themselves. In this paper, we describe a technique to calculate the 
dynamics of the thermal cloud using $N$-body simulations. 
Within this approach, a swarm of test 
particles is used to represent the evolution of the semiclassical 
phase-space density, while collisions are handled  
using a Monte Carlo sampling technique. The dynamics of the condensate
on the other hand is determined by numerically propagating the GP 
equation using a split-operator Fast Fourier Transform (FFT) method. 
Application of the method to the quadrupole
\cite{jackson02} and scissors \cite{jackson01b,jackson02b} modes has been discussed
elsewhere, and in both cases, good agreement with experiment
\cite{jin97,marago01} was found. Although an outline of the numerical
methods used was given in this earlier work, we give much more detail in
the present paper.

This paper is organized as follows. In Sec.\ \ref{sec:zgn} we briefly review 
the ZGN formalism, before discussing our numerical methods in Sec.\ 
\ref{sec:numer}. In Sec.\ \ref{sec:results} the Monte Carlo sampling is 
tested by comparison of equilibrium collision rates
against semi-analytic calculations. Landau and collisional damping rates for 
the monopole modes in spherical traps are also compared to previous 
theoretical treatments. We sum up and outline possible future research 
directions in the Conclusion.

\section{The ZGN formalism}
\label{sec:zgn}

We begin by reviewing the ZGN formalism, which was derived and 
discussed in detail in Ref.\ \cite{zaremba99}. For a Bose-condensed gas
one can decompose the second-quantized field operator 
$\hat{\psi}(\br,t)$ in the following manner
\begin{equation}
 \hat{\psi} (\br,t) = \Phi(\br,t)+\tilde{\psi} (\br,t),
\label{eq:decomp}
\end{equation} 
where the ensemble average $\Phi(\br,t)=\langle \hat{\psi}(\br,t)\rangle$
takes on a non-zero value due to Bose broken symmetry, and is identified
with the condensate wavefunction. The remaining field operator 
$\tilde{\psi} (\br,t)$ has a zero expectation value
and corresponds to the noncondensed component of
the cloud. The second-quantized Hamiltonian for the system is given by
\begin{eqnarray}
 \lefteqn{\hat{H} = \int {\rm d} {\br}\,\hat{\psi}^{\dagger} (\br) \left [
 -\frac{\hbar^2 \nabla^2}{2m} + U_{\rm ext} (\br) \right] \hat{\psi}
 (\br)} \nonumber \\
 & &\mbox{} + \frac{1}{2} \int {\rm d}\br \, {\rm d}\br'\,\hat{\psi}^{\dagger}
 (\br) \hat{\psi}^{\dagger} (\br') U_{\rm int} (\br,\br')
 \hat{\psi} (\br') \hat{\psi} (\br),
\label{eq:hamil}
\end{eqnarray}
where in most cases the trap is well approximated by a harmonic 
potential 
$U_{\rm ext} (\br) = m(\omega_x^2 x^2 +\omega_y^2 y^2
+\omega_z^2 z^2)/2$. We also assume a contact interaction,
$U_{\rm int}(\br,\br')=g\delta (\br-\br')$, with $g=4\pi\hbar^2 a/m$, where
$a$ is the $s$-wave scattering length and $m$ is the atomic mass. Using
$i\hbar \partial_t \hat{\psi} = [\hat{\psi},\hat{H}]$
with (\ref{eq:decomp}) and (\ref{eq:hamil}), one can derive coupled equations
of motion for the condensate and thermal cloud. In particular, the 
condensate order parameter evolves according to a generalized form of the 
GP equation
\begin{equation}
 i\hbar \frac{\partial}{\partial t} \Phi(\br,t) = \left (
 -\frac{\hbar^2 \nabla^2}{2m} + U_{\rm ext} (\br) + g[n_c (\br,t)+2\tilde{n}
 (\br,t)] - iR(\br,t) \right) \Phi(\br,t),
\label{eq:GP-gen}
\end{equation}
where $n_c(\br,t)=|\Phi(\br,t)|^2$ and $\tilde{n}(\br,t)=\langle 
\tilde{\psi}^{\dagger}(\br,t)\tilde{\psi}(\br,t) \rangle$ are the condensate
and noncondensate densities respectively. In arriving at this equation
we make the
Popov approximation whereby the so-called ``anomalous'' density, 
$\tilde{m}(\br,t)=\langle \tilde{\psi} (\br,t) \tilde{\psi}(\br,t)\rangle$, 
is neglected. This sidesteps problems associated with including this term,
such as ultraviolet divergences and an unphysical gap in the energy
spectrum at low momenta \cite{griffin96}. To go beyond this approximation
in a consistent manner requires a careful treatment of interparticle
collisions \cite{morgan00}, and is beyond the scope of the present work.
The source term $R(\br,t)$ is an important modification of the usual GP
equation as it allows the normalization of the wavefunction $\Phi$ 
to change with time. Physically this is due to collisions between 
condensate and noncondensate atoms which have the effect of transferring
atoms into or out of the condensate. The source term will be defined 
in terms of a collision integral later. 

It is convenient to describe the dynamics of the noncondensate in terms
of the Wigner operator \cite{zaremba99,kadanoff89}, which leads to the
definition of a phase-space distribution, $f(\bp,\br,t)$,
for the thermal excitations.
The equation of motion for the noncondensate can then be written as a 
kinetic equation 
\begin{equation}
 \frac{\partial}{\partial t} f (\bp,\br,t) + \frac{\bp}{m} \cdot \nabla
 f (\bp,\br,t) - \nabla U(\br,t) \cdot \nabla_{\bp} f (\bp,\br,t)
 = C_{12}[f] + C_{22} [f].
\label{eq:kinetic}
\end{equation}
\vfil\break
{\noindent In}
deriving this equation a number of approximations have been made, 
some of which have already been mentioned. 
Importantly, the excitations are assumed to be semiclassical within the
HF approximation; an excitation with momentum $\bp$ possesses an energy
$\epsilon=p^2/2m+U(\br,t)$, where
the effective potential $U(\br,t)=U_{\rm ext} (\br)+
2g[n_c (\br,t)+\tilde{n} (\br,t)]$ is composed of
the trap potential as well as mean fields from the condensate and 
the thermal cloud. The noncondensate density appearing in this
expression is given in terms of the distribution by
\begin{equation}
 \tilde{n} (\br,t) = \int \frac{{\rm d}\bp}{(2\pi \hbar)^3} 
f(\bp,\br,t)\,.
\label{eq:normdens}
\end{equation}
 
The terms on the right-hand side of (\ref{eq:kinetic}) are collision
integrals
that represent binary collisions between atoms. The $C_{22}$ term is 
familiar from the kinetic theory of a normal Bose gas, and corresponds
to the scattering of two atoms from initial to final thermal states. 
It is given by
\begin{eqnarray}
 C_{22} [f] &=& \frac{\sigma}{\pi h^3 m^2} \int {\rm d}\bp_2 {\rm d}\bp_3
 {\rm d}\bp_4 \delta(\bp+\bp_2-\bp_3-\bp_4) \nonumber \\
 && \qquad\qquad \times \delta(\epsilon+\epsilon_2
 -\epsilon_3-\epsilon_4) [(1+f)(1+f_2)f_3 f_4 - f f_2 (1+f_3)(1+f_4)].
\label{eq:c22}
\end{eqnarray}
where $f \equiv f (\bp, \br, t)$ and $f_i \equiv f (\bp_i, \br, t)$.
The total bosonic cross-section
is given by $\sigma=8\pi a^2$. The delta 
functions enforce momentum and energy conservation in the 
collision, while the factors $(1+f_i)$ account for Bose-enhancement of
the scattering.
The analogous $C_{12}$ collision integral corresponds to collisions 
that involve a condensate atom in either the initial or final states.
It is given by
\begin{eqnarray}
 C_{12} [f]&=&\frac{\sigma n_c}{\pi m^2} \int {\rm d}\bp_2 {\rm d}\bp_3 
 {\rm d}\bp_4 \delta(m\bv_c + \bp_2-\bp_3-\bp_4) 
 \delta(\epsilon_c+\epsilon_2-\epsilon_3-\epsilon_4) \nonumber \\
 &&\quad \times[\delta(\bp-\bp_2)-\delta(\bp-\bp_3)-\delta(\bp-\bp_4)]
 [(1+f_2)f_3 f_4 - f_2 (1+f_3)(1+f_4)],
\label{eq:c12}
\end{eqnarray}
where the local condensate velocity and energy are respectively given by
\begin{displaymath}
 \bv_c (\br,t) = \frac{\hbar}{2im|\Phi|^2} [\Phi^* \nabla 
 \Phi - \Phi\nabla \Phi^*],
\end{displaymath}
and
\begin{displaymath}
\epsilon_c = {1\over 2}mv_c^2 + \mu_c\,.
\end{displaymath}
Here, $\mu_c$ is the condensate chemical potential defined as
\begin{displaymath}
 \mu_c = -\frac{\hbar^2}{2m} \frac{\nabla^2 \sqrt{n_c}}{\sqrt{n_c}} 
 + U_{\rm ext} + gn_c + 2g \tilde{n}.
\end{displaymath}
If the condensed and noncondensed components are in local equilibrium,
the $C_{12}$ integral vanishes. Conversely, when the system is
perturbed from equilibrium the $C_{12}$ term acts to 
transfer atoms between the condensate and thermal cloud. These
collisions then define the source term in (\ref{eq:GP-gen}) according to
\begin{equation}
 R (\br,t) = \frac{\hbar}{2n_c} \int \frac{{\rm d}\bp}{(2\pi\hbar)^3}
 C_{12} [f].
\label{eq:rterm}
\end{equation}
The relative numbers of condensate and thermal particles will then 
adjust as a function of time until local equilibrium is re-established.

\section{Numerical methods}
\label{sec:numer}

In this section we describe the numerical methods used to 
solve the ZGN equations (\ref{eq:GP-gen})--(\ref{eq:rterm}) in the context
of a dynamical simulation. We first discuss the numerical methods 
used to solve
the GP and collisionless Boltzmann equations. Although these
are based on well-established techniques 
(see e.g.\ \cite{taha84,sanz-serna94}) we feel that our partly pedagogical
discussion will be useful for those trying to reproduce our simulations,
while highlighting the correspondence between the quantum and
classical dynamics of the system. We then move on to discuss treatment
of the $C_{22}$ and $C_{12}$ collision integrals by Monte Carlo sampling.
Finally, an overview of the simulations is provided, including a discussion
of how one calculates the equilibrium initial state of the system, as well as 
estimating the phase-space density in real time for use in evaluating
the collision integrals (\ref{eq:c22}) and (\ref{eq:c12}).

\subsection{The Gross-Pitaevskii equation}
\label{sec:GP-eq}

For the benefit of the following discussion we rewrite the 
GP equation (\ref{eq:GP-gen}) in the form
\begin{equation}
 i \hbar \frac{\partial}{\partial t} \Phi (t) = H (t) \Phi (t)\,.
\end{equation}
The time dependence of the Hamiltonian, $H(t) = T+V(t)$, arises from the
potential $V(t)$ which also includes the non-Hermitian source term 
$R({\bf r},t)$. In most of our simulations the time dependence is 
dominated by the nonlinear condensate potential and it is this term
which is the main source of numerical instabilities when the number of
condensate atoms is large. It is therefore important to develop a
numerical algorithm which is accurate even in this limit, and at the
same time, numerically efficient.

A formal solution of the above equation is given by
%
\begin{equation}
 \Phi (t+\Delta t) = U (t+\Delta t, t) \Phi (t)
\end{equation}
where the evolution operator $U$ has the expansion
\begin{equation}
U (t+\Delta t, t) = 1 + {1\over i\hbar} \int_t^{t+\Delta t} dt' H(t') +
{1\over (i\hbar)^2} \int_t^{t+\Delta t} \int_t^{t'} dt' dt'' H(t')
H(t'') + \cdots\,.
\end{equation}
Expanding the Hamiltonian as a Taylor series,
\begin{eqnarray}
H(t') &=& H(t) + {dH\over dt} (t'-t) + {1\over 2} {d^2 H \over
dt^2}(t'-t)^2 + \cdots \nonumber \\
&\equiv& \alpha + \beta (t'-t) + {1\over 2} \gamma (t'-t)^2 \cdots\,,
\label{Taylor}
\end{eqnarray}
we obtain
\begin{equation}
 U (t+\Delta t, t) = 1 + {\alpha \over i\hbar} \Delta t + {\beta \over
2i\hbar} (\Delta t)^2 -{\alpha^2 \over 2\hbar^2} (\Delta t)^2 
+ {\mathcal{O}} (\Delta t^3).
\label{Uexpansion}
\end{equation}
The lowest order exponential approximant to this expansion is
\begin{equation}
 U(t+\Delta t, t) \simeq {\rm e}^{-iH(t)\Delta t/\hbar} - 
\frac{i}{2\hbar}
 \frac{{\rm d}H}{{\rm d}t} (\Delta t)^2 + {\mathcal{O}} (\Delta t^3).
\label{eq:evol-op}
\end{equation}
The error of second order is shown explicitly. The first term on the
right hand side is of course exact for a time independent Hamiltonian 
but significant errors arise when the Hamiltonian is
time dependent. These errors can be
minimized by reducing the time step $\Delta t$, but at the expense of
increasing the computation time required to complete a simulation. Since
this imposes practical limits on the physical problems that can be
addressed, a more accurate approximant is desirable.

A higher order exponential approximant is provided by 
\begin{equation}
 U(t+\Delta t, t) \simeq {\rm e}^{-i(\alpha + {1\over 2} \beta \Delta
t)\Delta t/\hbar}.
\label{higher-order}
\end{equation}
A comparison with (\ref{Uexpansion}) indeed confirms that the error 
is 
${\mathcal{O}} (\Delta t^3)$. To this order of accuracy, we can make
use of (\ref{Taylor}) to estimate $\beta$ by reverse differencing,
\begin{equation}
\beta \simeq {H(t) - H(t-\Delta t) \over \Delta t}\,,
\end{equation}
and thus obtain
\begin{equation}
 U(t+\Delta t, t) \simeq {\rm e}^{-i\tilde H(t) \Delta t/\hbar}
 + {\mathcal{O}} (\Delta t^3)
\,,
\label{higher}
\end{equation}
where
\begin{equation}
\tilde H(t) = T + \tilde V(t)
\end{equation}
with
\begin{equation}
\tilde V(t) \equiv  {3V(t) - V(t-\Delta t) \over 2}\,.
\label{extrap}
\end{equation}
This is recognized as an approximation to the potential at time
$t+\Delta t/2$, the midpoint of the current time step, as obtained 
by a linear extrapolation from the
potential at times $t-\Delta t$ and $t$.

The implementation of (17-19) is very simple and costs only a small 
additional amount of memory to store the potential from the previous
time step. The actual numerical representation of the evolution operator
can be achieved by various methods. One popular approach is the
Crank-Nicholson method \cite{nr}, where finite-differencing Cayley's 
form for the operator leads to a set of linear equations
for the wavefunction at discrete grid-points in $\br$. The problem then
reduces to decomposition of a tridiagonal matrix at each time step and 
along each spatial dimension. 
In contrast, we favor a split-operator method, 
where a factorization of the exponential is effected by means of the
Baker-Campbell-Hausdorff (BCH) formula. One finds that
\begin{equation}
 {\rm e}^{-i\tilde H\Delta t/\hbar} = {\rm e}^{-i\tilde{V} \Delta 
t/2\hbar} 
 {\rm e}^{-iT \Delta t/\hbar} {\rm e}^{-i\tilde{V} \Delta t/2\hbar} + 
 {1\over 12} \left ( {\Delta t\over i\hbar}\right )^3
  \left[\left[T,\tilde{V}\right],
 \left(T +{\tilde{V}\over 2} \right)\right] + 
 {\mathcal{O}} (\Delta t^4).
 \label{eq:GP-ssop}
\end{equation}
The error generated by this approximation is of the same order as found
in (\ref{higher}). 
Applying the first term on the right hand side then evolves the wavefunction
to second order accuracy in $\Delta t$. In principle, higher order
schemes can be constructed by splitting into more elaborate combinations
of the $\tilde{V}$ and $T$ operators. However, to justify the effort, an
improved approximation for $\tilde H(t)$ is required. We have found that
second order accuracy is sufficient for most applications, although
difficulties do arise if the time scale of the simulations is 
exceedingly long.

The split-operator scheme (\ref{eq:GP-ssop}) is straightforward to implement
with a discrete grid in position space. The two potential steps are applied by
multiplying the wavefunction at each grid point by 
${\rm e}^{-i\tilde{V}\Delta t/2\hbar}$, 
while the kinetic term ${\rm e}^{-iT\Delta t/\hbar}$ is conveniently treated
in momentum space. The limiting step in the 
calculation is therefore the application of forward and inverse Fast Fourier
Transforms (FFTs) at each time step, but efficient FFT routines for 
arbitrary numbers of dimensions are readily available \cite{footnote1}. 
The dynamical evolution of the wavefunction can thus be followed over a series
of time-steps. Alternatively, stationary solutions of the time independent 
GP equation can be easily found by evolving the time-dependent equation
in imaginary time  $t \rightarrow -it$.

A typical application provides some indication of the relative merits 
of the higher order approximant
in (\ref{higher}) as opposed to the lower order scheme in
(\ref{eq:evol-op}). 
With the latter, one finds a monotonic increase in the energy 
expectation value with time. In simulations of
a collective mode this
effect would be apparent as a slow increase in the mode amplitude, which
is clearly undesirable when quantifying damping at finite temperatures.
More 
importantly, since the rate of increase scales with the mode energy, 
higher frequency excitations tend to build in amplitude more
rapidly. These excitations are initially generated at a low level by the
numerics; however over sufficiently long simulation times they 
eventually lead to instabilities in the wavefunction. These problems
are essentially eliminated with the higher order scheme. The stability
of the simulations is dramatically improved and the energy 
tends to oscillate with small amplitude about a constant value, 
rather than increasing
monotonically. The improved stability allows much larger
time-steps to be taken without compromising accuracy, leading to 
a considerable saving in computational effort.

\subsection{Collisionless particle evolution}
\label{sec:part-ev}

In this section we discuss solution of the collisionless Boltzmann 
equation ($C_{12}=C_{22}=0$) using $N$-body simulations. The effect of 
collisions is dealt with later.
Collisionless Boltzmann (or Vlasov) 
equations which include mean-field interactions arise
in many disparate fields, such as plasma physics, condensed matter
physics and astrophysics. Since the equation involves phase 
space variables in six dimensions, it is
generally very difficult to solve using standard methods for treating 
partial differential equations. An alternative approach used extensively
in the literature
is to represent the phase-space density $f(\bp,\br,t)$ by a cloud of discrete 
test particles \cite{hockney81}. The momentum and position of each particle 
in an external 
potential $U(\br,t)$ is then evolved according to Newton's equations. 
The phase space distribution for this situation is given by
\begin{equation}
 f (\bp,\br,t) \simeq  \frac{\tilde{N} h^3}{\tilde{N}_T} 
 \sum_{i=1}^{\tilde{N}_T} \delta [\br -\br_i (t)] 
\delta [\bp - \bp_i (t)]\,, 
\label{eq:discrete}
\end{equation}
where the weighting factor is fixed by the requirement that the phase-space 
distribution is normalized to the number of physical atoms,
$\tilde{N}$, with $ \tilde{N} h^3 = \int {\rm d}\br {\rm d}\bp f$.
By using a sufficiently large number of test 
particles, $\tilde{N}_T$, a reasonable approximation to the continuous
phase space distribution is obtained.
Note that the number of test and physical particles is not necessarily
equal. In fact, for a relatively small number of physical atoms 
($\tilde{N}\sim 10^4$) it is essential to simulate more test 
particles ($\tilde{N}_T > 10^5$) in order to minimize the effects of a
discrete particle description.
Conversely for large samples one can simulate fewer 
``superparticles'' so that the calculations are not too intensive.
 
The phase space variables are updated by advancing
the position and momentum of each particle at discrete time steps
$\Delta t$. This is not as trivial as one might naively expect. Conventional
integration schemes for ordinary differential equations, such as 
classical Runge-Kutta methods, can lead
to non-conservation of energy over long-time simulations when applied to
Hamiltonian systems. This results in
spurious damping or excitation of the system. In contrast, symplectic
integrators \cite{sanz-serna94,yoshida93} are used extensively in 
molecular dynamics (MD)
simulations since they possess several desirable properties, such
as conservation of phase-space volume and of energy over a long period
(as is required in autonomous Hamiltonian systems). We use a second-order 
symplectic integrator in our calculations,
which is the classical analogue of the split-operator method discussed earlier.
To show this, it is convenient to work within the Lie
formalism \cite{sanz-serna94}. Consider
the classical Hamiltonian for a single particle, $H_i = p_i^2 / 2m + V (\br_i)$.
The evolution of its phase-space coordinates 
$\bz_i=(\bp_i,\br_i)$ is then determined by the equation
\begin{equation}
 \frac{{\rm d}\bz_i}{{\rm d} t} = \{ \bz_i, H_i \}\equiv -i{\cal L}
\bz_i,
\end{equation}
where $\{ F,G \} = \sum_j \partial_{r_j} F \partial_{p_j} G - 
\partial_{p_j} F \partial_{r_j} G$ is the Poisson bracket and ${\cal L}$
is the Liouville operator \cite{prigogine}. One can then write
\begin{equation}
 \bz (t+\Delta t) = {\rm e}^{-i{\cal L}  \Delta t} \bz (t).
\end{equation}
Splitting the Hamiltonian into potential and kinetic terms, 
$H_i = T (\bp_i) + V (\br_i)$, the BCH formula can be used again
to show that \cite{yoshida93}
\begin{equation}
{\rm e}^{-i{\cal L} \Delta t} =  {\rm e}^{-i{\cal L}_T \Delta t/2} {\rm
e}^{-i{\cal L}_V \Delta t} {\rm e}^{-i{\cal L}_T \Delta t/2}
 - \frac{(\Delta t)^3}{12} \left \{ \left \{ T,V \right \},
 \left( V + \frac{T}{2} \right ) \right \} 
 + {\mathcal{O}} (\Delta t^4).
\label{eq:class-ssop}
\end{equation}
One now sees the analogy with the quantum operator (\ref{eq:GP-ssop}),
where both conserve energy to order $(\Delta t)^2$. The effect of the
classical operator (\ref{eq:class-ssop}) in the simulations is to update the
particle positions and velocities in three steps
\begin{eqnarray}
 & \tilde{\br}_i = \br_i (t) + {1\over 2}\Delta t \bv_i (t), &
 \nonumber \\
 & \bv_i (t+\Delta t) = \bv_i (t) - m^{-1} \Delta t \nabla V 
 (\tilde{\br}_i), & \nonumber \\
 & \br_i (t+\Delta t) = \tilde{\br}_i + {1\over 2} \Delta t \bv_i 
 (t+\Delta t). &
\end{eqnarray}
By analogy with (\ref{extrap}), $V$ should be the midpoint value
of the potential, $\tilde V(t)$, when it is time-dependent. In our
simulations, $V$ is the effective potential $U(\br,t) = U_{\rm ext}(\br)
+2gn(\br,t)$ felt by the thermal atoms, where $n=n_c + \tilde n$ is the
total density.
 
\subsection{Thermal Cloud Potential}
\label{sec:potential}

The effective potential $U$ is determined
self-consistently as the system evolves in time, and includes
the condensate mean field $2gn_c(\br,t)$ and the mean field generated by
the thermal cloud $2g\tilde n(\br,t)$.
The latter is in general much weaker than the condensate mean field 
due to the larger spatial extent (and therefore
lower density) of the thermal cloud. Nevertheless, it is important 
to include this term in order to ensure the conservation of the total
energy of the system. In addition, from the perspective of
the condensate, the noncondensate mean field is necessary in order to
account for the temperature-dependent damping and frequency shifts of 
condensate collective modes.

Although the calculation of the condensate mean field is
straightforward, the use of discrete particles with a contact
interatomic potential creates a problem in determining
the noncondensate mean field. Taken literally, the mean field 
consists of a series of delta peaks
\begin{equation}
 \tilde{U}_T (\br, t) = 2g \frac{\tilde{N}}{N_T} \sum_i^{\tilde{N}_T}
 \delta (\br - \br_i) \equiv 2g \tilde n_T(\br,t)\, .
\end{equation}
This expression clearly cannot be used as it is to generate the forces
acting on the test particles that are required in the MD simulation.  
Rather, the density $\tilde n_T(\br,t)$ must be replaced by
a smooth and differentiable thermal cloud density
and some smoothening operation is therefore needed.
A possible first step might be to
divide space into cells and to determine the mean density within each
cell by binning the test particles appropriately. However, this binning
procedure generates spatial discontinuities on the scale of the 3D grid
being used that would still have to be smoothed out in some way. In
addition, temporal discontinuities arise as particles migrate from one
cell to another. These
temporal fluctuations are of course spurious since they depend on the
number of test particles and decrease in relative
amplitude as this number is increased. It is apparent that the binned
density has some undesirable properties associated with the
statistical fluctuations in the number and positions of particles in
each cell.

As an alternative to this binning procedure, we generate a smooth 
thermal cloud density
by performing a convolution with
a sampling (or smoothening) function $S(\br)$ which is normalized to
unity. In particular, we define
\begin{displaymath}
\tilde U_S(\br,t) \equiv  \int {\rm d} \br' S (\br-\br') \tilde{U}_T(\br', t) = 
 2g \frac{\tilde{N}}{\tilde{N}_T} \sum_i^{\tilde{N}_T} S (\br -
\br_i)\,,
\end{displaymath}
where we
choose $S(\br) \sim e^{-r^2/\eta^2}$, i.e.\ an isotropic Gaussian
sampling function of width $\eta$. Since $\nabla S|_{\br=0}$,
no force is exerted by a particle on itself and the sum can extend
over all particles in the ensemble. 
Ideally, the width of $S(\br)$ should be small compared
to the curvature of the noncondensate density. If, at the same time,
the number of particles contributing to the sum at a given position 
$\br$ is large, it is clear that the sampled potential will
be relatively smooth. Note that the smoothening operation is equivalent
to assuming a finite-ranged interatomic potential.
 
The sampled potential (or its gradient) 
is needed at the position of each test particle and
at the mesh points on which the condensate wavefunction is defined.
However a direct summation for all points would be prohibitive. 
We therefore
proceed by making use of a FFT. First, each particle in the ensemble is
assigned to points on the 3D Cartesian grid using a cloud-in-cell 
method \cite{hockney81}.  This is most readily explained in 1D: consider
a particle at position $x$, between two grid points at $x_k$ and
$x_{k+1}$. The particle is assigned to both points with weightings
$(1-\alpha)$ and $\alpha$ respectively, where $\alpha=(x - x_k)/
(x_{k+1}-x_k)$.  This can be viewed as a
more sophisticated binning procedure in that it takes into account the
actual positions of particles within the cells.
The generalization to 3D is straightforward, where in
this case the particle is assigned to the eight points which define the
unit cell containing the particle. We then convolve the cloud-in-cell
density with the sampling function by Fourier transforming it and then 
multiplying it by the analytic FT of the sampling function. 
An inverse FFT then generates the sampled potential on the 3D grid. 
This potential is used directly in the GP evolution, while the 
forces on the test particles are obtained by taking a numerical 
derivative and interpolating to the positions of the particles.

\begin{figure}
\centering
\psfig{file=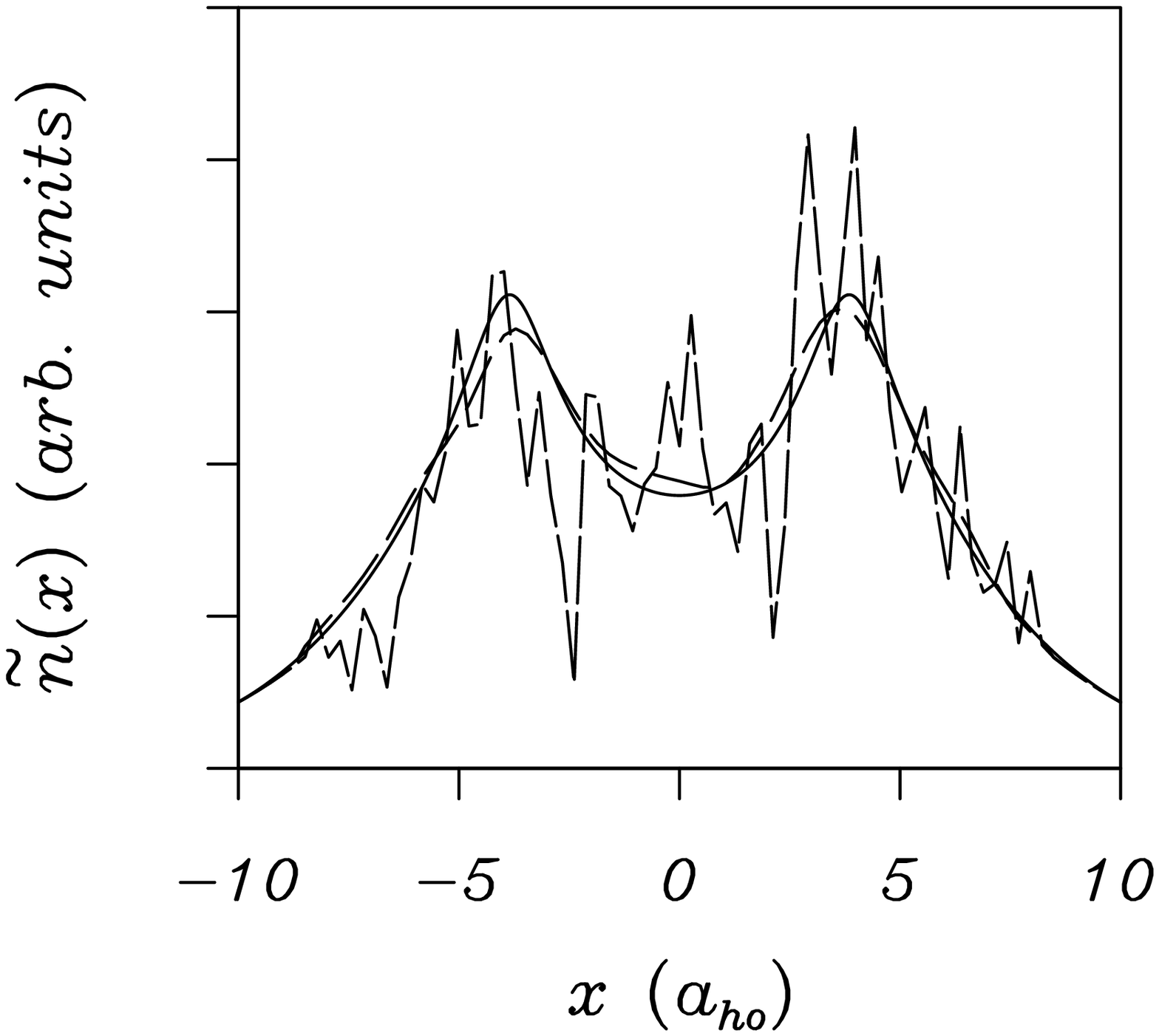, scale=0.53, bbllx=80, bblly = 115,
 bburx=570, bbury=560} 
 \caption{Equilibrium noncondensate density against position, along a line
 through the center of an isotropic trap with frequency 
 $\omega_0 = 2\pi \times 187\, {\rm Hz}$. The system consists of a total 
 of $N_{\rm tot}=5\times 10^4$ $^{87}{\rm Rb}$ atoms at a temperature of
 $T=250\,{\rm nK}$. The critical temperature for an equivalent ideal gas 
 would be $T_c^0=310.6\,{\rm nK}$. $\tilde{N}_T = 4.0\times 10^5$ test 
 particles are sampled according to the actual equilibrium density (solid
 line). The fluctuating dashed line is a result of binning particles 
 using a cloud-in-cell method, while the smooth dashed line shows the effect
 of convolving the cloud-in-cell density with a Gaussian.}
 \label{fig:density} 
\end{figure}

This overall scheme is illustrated in Fig.~\ref{fig:density}. The solid line shows the
equilibrium thermal cloud density along a line through the center of an
isotropic trap with trapping frequency $\omega_0 = 2\pi \times 
187\,{\rm Hz}$,
a system we study in more detail later. The trap contains a total of 
$N_{\rm tot}= 5 \times 10^4$ $^{87}$Rb atoms, and at a temperature of 
$T= 250\,{\rm nK}$
there are $\tilde N \simeq 4.0 \times 10^4$ thermal atoms. The rapidly
fluctuating dashed line is the density along this line produced by the
cloud-in-cell method using a thermal distribution of $\tilde N_T \simeq
4.0\times 10^5$ test particles, that is, ten times the actual number of
thermal atoms. The effect of statistical fluctuations is clearly
evident. Finally, the smooth dashed line is the result of the
convolution using a width parameter of $\eta \simeq 0.76\,a_{ho}$, where
$a_{ho} = (\hbar/m\omega_0)^{1/2} \simeq 7.9 \times 10^{-7}\,{\rm m}$ is the 
harmonic oscillator length
for the trap being considered. (For comparison, the mesh size is $\Delta
x \simeq 0.27\,a_{ho}$.) It should be noted that the dramatic
smoothening of the density achieved is partly a consequence of
performing a full 3D convolution; a 1D convolution of the cloud-in-cell
density with the same width parameter would not reduce the amplitude of
the spatial fluctuations to the same degree.
Finally, we compare the convolved density to the actual
equilibrium density. Apart from differences due to the statistical
sampling of test particles, one can see that the peaks in the thermal
cloud density at the edges of the condensate are slightly broader in the
convolved density, as would be expected. However, the differences are
minor and do not affect the dynamics of the system significantly. For
consistency, the $n_c$ term appearing in $U(\br,t)$ is also convolved.

\subsection{Collisions}   
 
The methods outlined so far allow one to follow the condensate wavefunction 
and trajectories of the atoms subject to a time-dependent potential, so long
as the system is in the collisionless regime. However in general the 
collisional terms
in the Boltzmann equation will be nonzero $C_{22}\neq 0$, $C_{12}\neq 0$.
In other words, during each time step there is a certain probability
that a given test particle will collide with another thermal atom or with 
the condensate. If the typical collision timescale $\tau$ is such
that $ \tau  \gg \Delta t$,
one can treat the free particle evolution and collisions separately.
Each particle's trajectory is first followed using the methods discussed
in the previous section, and the possibility of collisions occurring
is then considered at the end of the time step.
Probabilities for either $C_{22}$ or $C_{12}$
collisions are calculated in a way which is consistent with 
a Monte Carlo sampling of the collision integrals, as discussed below. 

\subsubsection{$C_{22}$ collisions}

We first give details for the $C_{22}$ integral (\ref{eq:c22}), which 
physically corresponds to scattering of two thermal particles into two final 
thermal states. Hence the process conserves the number of thermal atoms, 
$ \int {\rm d}\bp/(2\pi\hbar)^3 C_{22} = 0$. We are interested in the 
mean collision rate at a point $\br$ (as defined in Appendix A), 
which is given by 
\begin{eqnarray}
 \Gamma_{22}^{\rm out} &=& \frac{\sigma}{\pi h^6 m^2} \int {\rm d} 
 \bp_1 f_1 \int {\rm d} \bp_2 f_2 \int {\rm d} \bp_3 \int {\rm d} \bp_4
\delta (\bp_1+\bp_2-\bp_3-\bp_4) 
 \nonumber \\
&&\qquad\qquad\qquad\qquad \times
\delta(\epsilon_1+\epsilon_2-\epsilon_3
 -\epsilon_4) (1+f_3)(1+f_4).
\label{eq:22-rate}
\end{eqnarray} 
For our purposes it is convenient to express the integral in terms of
new momentum variables ($\bp_0$, $\bp_0'$) and ($\bp'$, $\bp''$):
$\bp_{1,2}=(\bp_0 \pm \bp')/\sqrt{2}$ and $\bp_{3,4}=(\bp_0'\pm
\bp'')/\sqrt{2}$. $\bp_0$ and $\bp'$ are proportional to the
center-of-mass and relative momenta, respectively, of the incoming 1 and
2 particles.
By implicity assuming energy and momentum conservation
($\bp_0=\bp_0'$, $p'=p''$) one can rewrite (\ref{eq:22-rate}) in the 
simplified form
\begin{equation}
 \Gamma_{22}^{\rm out} = \int \frac{{\rm d}\bp_1}{(2\pi\hbar)^3} f_1
 \int \frac{{\rm d}\bp_2}{(2\pi\hbar)^3} f_2 \int \frac{{\rm d}\Omega}{4\pi}
 \sigma |\bv_1-\bv_2| (1+f_3)(1+f_4)\,,
\label{gamma22_out}
\end{equation} 
where $\bp_{3,4}=(\bp_0\pm p' \hat {\bf u}(\Omega))/\sqrt{2}$, with
$\hat {\bf u}(\Omega)$ a unit vector in a direction 
specified by the solid angle $\Omega$.
Calculation of the rate therefore involves integrals over all possible 
initial states and all scattering angles $\Omega$. 
In the equilibrium situation,
this rate defines a local mean collision time $\tau^0_{22}$ according to
\begin{equation}
 \Gamma_{22}^{0} \equiv {\tilde n_0 \over \tau_{22}^0}\,,
\label{eq:gam022}
\end{equation}
where $\tilde n_0(\br)$ is the equilibrium thermal cloud density.
As shown in \cite{jackson02b}, $1/\tau_{22}^0$ below $T_c$ is a strong
function of position for a trapped Bose gas and is peaked at the edge 
of the condensate. In the classical (i.e.\ Maxwell-Boltzmann) limit,
$1/\tau^0_{22}$ reduces to $\sqrt{2}\sigma v_{th} \tilde n_0$, with
$v_{th} = (8kT/\pi m)^{1/2}$.

To relate this to collision probabilities for individual atoms in our 
simulations requires sampling of the integral using
a rejection method as discussed in detail in Appendix A\cite{nr,wu97}. 
At each time step atoms are first 
binned into cells of volume $\Delta^3r$ according to their position. 
The atoms within each
cell are then paired at random, and a probability for a pair $(ij)$ to
collide in the time step $\Delta t$ is assigned according to
\begin{equation}
 P_{ij}^{22} = \tilde{n} \sigma |\bv_i-\bv_j| \int \frac{{\rm d}\Omega}{4\pi}
 (1+f_3) (1+f_4) \Delta t\,.
\label{avgprob}
\end{equation}
The integral over $\Omega$ can be evaluated by averaging over a sample of 
randomly selected final states which are obtained by choosing 
uniformly-distributed random 
values for the scattering variables $\cos\theta$ and $\phi$. However, 
in simulating the collision process, the velocities of the incoming 
particles must 
actually change to a specific, but random, pair of final velocities.
These velocities lie on a sphere centered at $(\bv_1+\bv_2)/2$ with
a radius $|\bv_1-\bv_2|/2$ and can be chosen by randomly selecting the
scattering angle $\Omega_R$. The appropriate collision probability for
this event is then
\begin{equation}
 P_{ij}^{22} = \tilde{n} \sigma |\bv_i - \bv_j| (1+f_3^{\Omega_R})
 (1+f_4^{\Omega_R}) \Delta t.
\label{eq:22-prob}
\end{equation}
This probability depends upon the phase space densities of the final
states, $f_3^{\Omega_R}$, $f_4^{\Omega_R}$, reflecting Bose statistics.
If this single scattering probability is averaged over a random 
distribution of
scattering angles $\Omega_R$ we recover the average probability defined
in (\ref{avgprob}). 

The simulation of $C_{22}$ collisions thus proceeds as follows. A pair
of test particles $(ij)$ in a given cell is chosen at random. Whether a
collision of this pair occurs is then tested by
comparing $P_{ij}^{22}$ to a random number $X^{22}$ uniformly 
distributed between 0 and 1. If $X^{22}<P_{ij}^{22}$ the collision is 
accepted, and
the velocities of the test particles are updated accordingly. If 
$X^{22}>P_{ij}^{22}$, no collision occurs and the velocities of the
colliding pair are unchanged. In either case, another pair is randomly
selected and the procedure is repeated for all pairs in each cell of 
the sample.

\subsubsection{$C_{12}$ collisions}   

The $C_{12}$ collisions are treated in a similar manner to $C_{22}$. 
The key difference here is that one of the collision partners is a
condensate atom in a definite state, and it is necessary to distinguish
the collisional processes which
either transfer an atom into or out of the condensate. For example, 
the ``out''
collision rate as defined in (\ref{Agamout}) is given by
\begin{equation}
 \Gamma_{12}^{\rm out} = \frac{\sigma n_c}{\pi m^2 h^3} \int {\rm d}\bp_2
 {\rm d}\bp_3 {\rm d}\bp_4 \delta (\bp_c+\bp_2-\bp_3-\bp_4)\delta(\epsilon_c
 +\epsilon_2-\epsilon_3-\epsilon_4) f_2 (1+f_3)(1 +f_4).
\label{eq:12out-rate}
\end{equation}
This represents scattering of a thermal atom from the condensate
to produce two thermal atoms. The reverse process gives the ``in''
collision rate defined in (\ref{Agamin}):
\begin{equation}
 \Gamma_{12}^{\rm in} = \frac{\sigma n_c}{\pi m^2 h^3} \int {\rm d}\bp_2
 {\rm d}\bp_3 {\rm d}\bp_4 \delta (\bp_c+\bp_3-\bp_2-\bp_4)\delta(\epsilon_c
 +\epsilon_3-\epsilon_2-\epsilon_4) f_2 (1+f_3) f_4.
\label{eq:12in-rate}
\end{equation}
In obtaining (\ref{eq:12in-rate}) we have interchanged the 2 and 3
labels in order to define an integral having the same $f_2$ weighting
factor as in (\ref{eq:12out-rate}). These two integrals give the true
``in" and ``out" collision rates. However in the simulations,
it is useful to drop the cubic terms $f_2 f_3 f_4$ 
which formally cancel exactly between the ``in" and ``out" rates. Since
these two rates are evaluated differently as explained below, this
cancellation will not be numerically precise, and it is therefore
preferable to eliminate the cubic terms from the calculation of
collision probabilities. In the following, we denote the rates with the
cubic terms removed by ${\overline \Gamma}_{12}^{\rm in(out)}$. 
Dropping these terms of course does not change the {\it net}
rate of transfer from the condensate to the thermal cloud that actually
takes place.

The ``out'' term can be reduced by transforming the momentum variables
as before, with the result
\begin{equation}
{\overline \Gamma}_{12}^{\rm out} = \int \frac{{\rm d}\bp_2}{(2\pi\hbar)^3} f_2 n_c
 \sigma v_r^{\rm out} \int \frac{{\rm d}\Omega}{4\pi} (1+f_3+f_4),
\label{eq:gam012}
\end{equation}
where $v_r^{\rm out}= \sqrt{|\bv_c-\bv_2|^2-4gn_c/m}$ is the relative 
velocity of the
initial states, corrected to account for
energy conservation (locally, the mean field energy of a thermal atom 
is higher than that of a condensate atom by an amount $gn_c$). 
Now, if we consider each atom in the distribution $f_2$ in 
turn, the probability for collision with the condensate is given by
\begin{equation}
P_i^{\rm out} = n_c \sigma v_r^{\rm out} (1+f_3^{\Omega_R}+
f_4^{\Omega_R}) \Delta t\,.
\label{eq:out-prob}
\end{equation}
In this case, the final thermal atom velocities $\bv_3$, $\bv_4$ lie 
on a sphere of radius $v_r^{\rm out}/2$ centered on $(\bv_c+\bv_2)/2$, 
with a random scattering angle $\Omega_R$.

``In'' collisions involve scattering of two thermal atoms to produce a 
condensate and a thermal atom. In the context of (\ref{eq:12in-rate}),
the incoming atoms are labelled 2 and 4, and the outgoing thermal
atom is labelled 3. Energy-momentum conservation in (\ref{eq:12in-rate})
dictates the condition $(\bp_c-\bp_2)\cdot(\bp_c-\bp_4)=mgn_c$.
Thus, unlike the case of $C_{22}$ collisions, one cannot arbitrarily
select a pair of 2 and 4 atoms from the sample since this condition will
in general
be violated and the collision cannot occur. To proceed, we perform the
integrations involving the delta functions in (\ref{eq:12in-rate}) to
obtain
\begin{equation}
{\overline \Gamma}_{12}^{\rm in} = \int \frac{{\rm d}\bp_2}{(2\pi\hbar)^3} f_2 
 \frac{n_c \sigma}{\pi v_r^{\rm in}} \int {\rm d}\tilde{\mathbf{v}} f_4,
\end{equation}                      
where $\bv_r^{\rm in}\equiv \bv_2-\bv_c$ is the velocity of
thermal atom 2 relative to the local condensate velocity. 
The second integral is a
two-dimensional integral over a velocity vector $\tilde{\mathbf{v}}$ 
which is in a plane normal to $\bv_r^{\rm in}$. The
velocity of the other incoming thermal atom, particle 4, is given by
\begin{displaymath}
 \bv_4 = \bv_c + \tilde{\bv} + \frac{gn_c}{mv_r^{\rm in}}\hat 
\bv_r^{\rm in}\,,
\end{displaymath}
while the velocity of the outgoing thermal atom is
\begin{displaymath}
 \bv_3 = \bv_2 + \tilde{\bv} + \frac{gn_c}{mv_r^{\rm in}}\hat 
\bv_r^{\rm in}\,.
\end{displaymath}

In the simulation one
considers each thermal atom in the distribution $f_2$ in turn, 
then randomly selects two 
numbers that define the vector $\tilde{\mathbf{v}}=\tilde{\mathbf{v}}_R$
within a plane of area ${\mathcal{A}}_v$. 
The collision probability is then given by
\begin{equation}
 P_i^{\rm in}= \frac{n_c \sigma {\mathcal{A}}_v}{\pi v_r^{\rm in}} 
 f_4^{\tilde{\bv}_R} \Delta t\,.
\label{eq:in-prob}
\end{equation}
Note that the area ${\mathcal{A}}_v$ appears in this expression, which at
first sight is disconcerting since it is an arbitrary number entering
as a simulation parameter. However, we find that the total rate is
largely independent of this area so long as the plane  
completely samples the occupied regions of phase space. We show results
confirming this statement in the following section.
  
This analysis yields probabilities for a particular atom to undergo 
``out" or ``in'' collisions. To decide whether either event takes 
place, another random number $0<X^{12}<1$ is chosen. If 
$X^{12}<P_i^{\rm out}$ then an ``out''
collision is accepted; the incoming thermal atom is removed from the
ensemble of test particles and two new thermal atoms are created. 
On the other hand,  
if $P_i^{\rm out}<X^{12}<P_i^{\rm out}+P_i^{\rm in}$, then an ``in'' 
collision takes place and atom 2 is removed from the thermal sample.
In addition, a second test particle, atom 4, is removed and a new
thermal atom, atom 3, is created. In practice, it is exceedingly
unlikely that a test particle will exist that will precisely match the
required phase-space coordinates of particle 4. We therefore search for
a test particle in neighboring phase-space cells and remove this
particle if one is found. 
This can be justified by remembering that we are only interested in 
describing the evolution in phase space in a statistical way---it is 
misleading to think of a direct correspondence between the test 
particles and physical atoms. If no test particle exists in the
vicinity of $\bv_4$, the local phase-space density $f_4$, and hence
$P_i^{\rm in}$,  will be zero and the ``in" collision is precluded from
occuring in any case.

The above procedure leads to a change in the number of atoms in the thermal
cloud. In order to conserve the total particle number the GP equation 
(\ref{eq:GP-gen}) is propagated with the $R$-term  which changes the 
normalization of the wavefunction and hence the condensate number. This 
quantity can be evaluated from the Monte Carlo process decribed above by
summing probabilities for particles around each grid point $\br_{jkl}$
using (\ref{eq:rterm}), i.e.,
\begin{equation}
 R (\br_{jkl},t) = \frac{\hbar}{2n_c \Delta t} \sum_i (P_i^{\rm out} - 
 P_i^{\rm in}).
\label{eq:R-calc}
\end{equation}
In practice this assignment to grid points is performed with a 
cloud-in-cell approach similar to the one described earlier. 
Of course the normalization of the condensate wavefunction varies
continuously as opposed to the variation of the thermal atom number
which changes by discrete jumps. Nevertheless, one
can show that the subsequent change in the condensate normalization 
is consistent with the addition or removal of atoms from the 
thermal cloud, so that the total particle number, $N_{\rm tot}$,
is conserved within statistical fluctuations ($\sim\sqrt{N_{\rm tot}}$).

\subsection{Overview}  
\label{sec:overv}

So far we have described various aspects of the numerical scheme. The 
aim of this subsection is to tie these disparate elements together with
an overview of the simulation procedure as a whole. One of the main
applications of our approach is to the study of small amplitude
collective oscillations around the equilibrium state. 
The first requirement of such a calculation is therefore the
self-consistent determination of the equilibrium thermal cloud 
distribution and condensate wavefunction.  Since the
thermal excitations are treated semiclassically, the thermal cloud is
described by the equilibrium Bose distribution
\begin{equation}
 f_0 (\bp,\br) = \frac{1}{z^{-1} e^{\beta p^2/2m}-1},
\label{eq:bose-dist}
\end{equation}
where $z(\br)=\exp \{\beta[\mu_c-U(\br)] \}$ is the local fugacity and 
$\beta \equiv 1/k_B T$. It is straightforward to show that both the 
$C_{12}$ and $C_{12}$ collision integrals vanish in this case.
The noncondensate density profile can be evaluated from 
(\ref{eq:normdens}) and (\ref{eq:bose-dist}) to yield
\begin{equation}
 \tilde{n}_0 (\br) = \frac{1}{\Lambda^3} g_{3/2} (z)
\label{eq:normdens2}
\end{equation}
where $\Lambda = (2\pi \hbar^2/mk_B T)^{1/2}$ is the thermal de Broglie
wavelength. The equilibrium condensate wavefunction is obtained as the
stationary solution of (\ref{eq:GP-gen}), with $R=0$, and the
corresponding eigenvalue defines the equilibrium chemical potential
$\mu_c$. Since the condensate and thermal cloud are coupled by mean
fields, the two components have to be determined self-consistently using
an iterative procedure. Details of this have been given by several 
authors (see e.g.\ \cite{zaremba99} or \cite{williams01a}) and will 
not be repeated here. 

To represent the thermal cloud in the simulations, an ensemble of test
particles must be defined. In the case of an equilibrium situation, this
ensemble should have a phase-space distribution which is consistent with
the Bose equilibrium distribution in (\ref{eq:bose-dist}). This can be 
achieved using the following rejection algorithm \cite{nr}. First, we
distribute particles in position space according to the density $\tilde
n(\br)$. To do this,
we select three random numbers uniformly distributed between
$-r_{\rm max}$ and $r_{\rm max}$, defining Cartesian coordinates, 
$\br_i$, of a particle  
in the occupied region of position space. A further uniform deviate is
then chosen from $R_i^1 \in [0,\tilde{n}_{\rm max}]$, where 
$\tilde{n}_{\rm max} \geq {\rm max} \{ \tilde{n} (\br) \}$, and compared
to the density at that point $\tilde{n} (\br_i)$. If $R_i^1 > \tilde{n} 
(\br_i)$,
the particle is discarded and another set of position coordinates 
selected. Otherwise, if $R_i^1 < \tilde{n} (\br_i)$, the particle is
accepted and one proceeds to specify its momentum by choosing
another random number $p_i \in [0,p_{\rm max}]$.
A random number $R_i^2 \in [0,f_{\rm max}]$ (where $f_{\rm max} \geq
z(\br_i)/[1-z(\br_i)]$, with $z(\br_i)$ the local fugacity) is 
compared to $f (p_i ,\br_i)$ to decide 
whether the momentum is accepted or rejected. In the case of rejection 
another $p_i$ is chosen, while if accepted two random angles are 
selected $\phi \in [0,2\pi]$, $\cos \theta \in [-1,1]$, which in turn 
define the momentum vector $\bp_i$. This procedure is repeated until
$\tilde{N}_T$ test particles in the ensemble are accumulated. 
Note that we have exploited the 
spherical symmetry of the equilibrium distribution in momentum space. 
In principle, a similar method can be applied to position space if the 
trap is spherically or cylindrically symmetric. 

A dynamical simulation can be initiated in one of two ways. Either an
appropriate nonequilibrium initial state is specified, or the system is
dynamically excited with the application of an external perturbation.
The latter parallels the procedure used experimentally to study
small amplitude collective excitations, and usually amounts to some
parametric manipulation of the trapping potential. Although this might
be the preferred approach, it is not always the most appropriate, 
especially when the excitation phase requires a prohibitively long
simulation time. It is then more convenient
to impose the perturbation on the initial state itself.
Here we are guided by the nature and symmetry of the collective mode
being studied, as well as information gleaned from earlier
calculations such as those based on the Thomas-Fermi approximation.
For example, the nature of the density fluctuation or velocity field 
associated with the mode might be known and it is then advantageous to
use this information in defining the initial state. A good example of
this is the breathing, or monopole, mode in an isotropic trap.
In this case the TF mode has a velocity field $\bv = a \br$. 
To impose this
velocity on the condensate one can simply multiply the ground state
wavefunction by a phase factor $\exp(imar^2/2\hbar)$. In the case of 
the thermal cloud, the same velocity field can be imposed by adding
$a\br_i$ to the velocity of the $i$-th particle in the equilibrium
ensemble. This procedure will predominantly excite the lowest monopole
oscillation. Although higher lying modes might also be mixed in to some
extent, they have different frequencies and can usually be separated
from the dominant mode when analyzing the dynamics.  

Returning to the simulation procedure itself, the condensate
wavefunction and thermal atom phase-space coordinates are updated in
each time step $\Delta t$ according to the prescription detailed in 
Sec.\ \ref{sec:part-ev}. Then, before treating collisions the thermal 
atoms are assigned to cells in position space. These are used for 
selecting pairs for $C_{22}$ collisions, as well as being further 
subdivided into momentum space elements in order to estimate the 
phase space density $f(\bp,\br)$ for calculating 
collision probabilities. Since collisions are
treated one cell at a time, the
phase space density only needs to be calculated and stored for
one particular cell. The $C_{12}$ and $C_{22}$ collisions are then
treated using the Monte Carlo scheme described earlier and the momenta
and number of thermal atoms (test particles) are updated.
Repeating for all of the cells yields the quantity $R$ from  
(\ref{eq:R-calc}) which, when used in the GP propagation 
(Sec.\ \ref{sec:GP-eq}), continuously evolves the number of atoms in 
the condensate.  For numerical accuracy the positional cells 
should enclose regions of almost constant thermal density and fugacity,
and are most conveniently treated using a spatial grid which reflects 
the (elliptical) geometry of the cloud. The momentum elements
in contrast lie on a Cartesian grid, where a cloud-in-cell method allows
one to minimize statistical fluctuations while retaining a fine grid for
precision.

\section{Results}   
\label{sec:results}

\subsection{Equilibrium collision rates}
\label{sec:equil}

Our first calculations are not simulations as such, but are instead 
checks of the Monte Carlo sampling technique we use to evaluate the 
$C_{12}$ and $C_{22}$ collision rates in real time. The physical
situation we consider corresponds to the one discussed at the end of
Sec.~\ref{sec:potential}, namely $5\times 10^4$ $^{87}$Rb atoms at 250 nK
in an isotropic trap. The equilibrium $C_{22}$ collision rate
$\Gamma_{22}^0$ can be evaluated numerically
directly from the expression in (\ref{gamma22_out})
using the equilibrium distribution function (\ref{eq:bose-dist}). The
result as a function of the radial coordinate $r$ is shown as the solid
line in Fig.~2. The equilibrium $C_{12}$ collision rates can also be
calculated using the equilibrium distribution (\ref{eq:bose-dist}) and
equilibrium condensate density $n_c(r)$ in (\ref{eq:12out-rate}) or 
(\ref{eq:12in-rate}). The ``in" and ``out'' rates are in fact equal to 
each other in equilibrium and will be denoted $\bar \Gamma_{12}^0$
(recall that these rates are calculated ignoring the
cubic terms in the full expression). The result of the calculation as
a function of $r$ is shown as the solid line in Fig.~3.
One sees that both the $C_{12}$ and $C_{22}$ collision rates exhibit a
maximum near to the condensate surface, where the fugacity $z$ 
approaches unity and the equilibrium Bose distribution is strongly 
peaked at $\bp=0$. However, in the case of
$C_{22}$ collisions, the tail of the distribution
decays more slowly since the thermal cloud density extends out to larger
radii than the condensate. 

\begin{figure}
\centering
\psfig{file=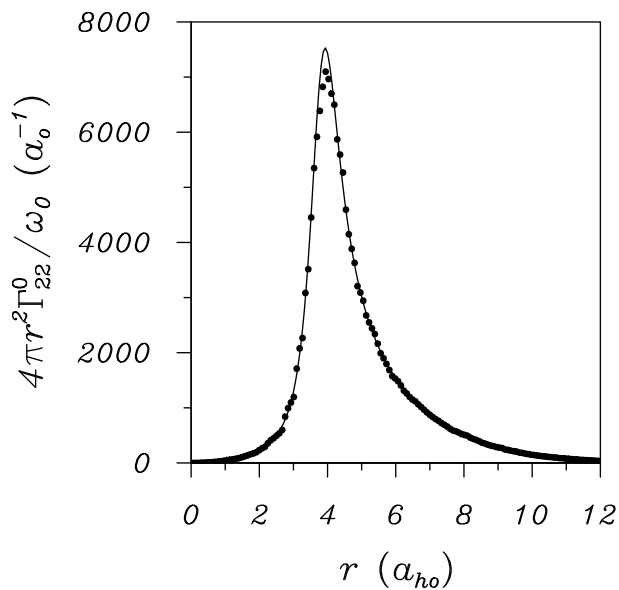, scale=0.42, bbllx=50, bblly = 70,
 bburx=600, bbury=610} 
 \caption{The $\Gamma_{22}^0$ collision rate as a function of position, for
  an equilibrium distribution and the same parameters as Fig.\ 
  \ref{fig:density}. The solid line shows the 
  result of a direct evaluation of (\ref{eq:gam022}), while the points 
  plot the results of a Monte Carlo evaluation (\ref{avgprob}).}
 \label{fig:eqm-c22} 
\end{figure}

\begin{figure}
\centering
\psfig{file=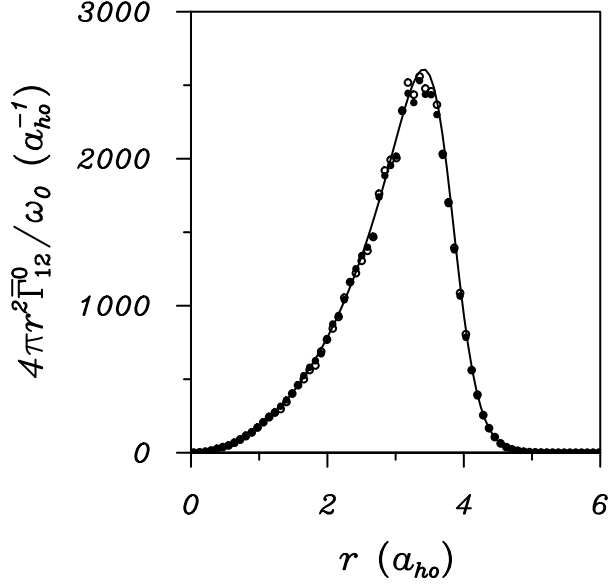, scale=0.42, bbllx = 50, bblly = 70,
 bburx=595, bbury=610}
 \caption{Same parameters as in Fig.\ \ref{fig:density}, for $
 \overline{\Gamma}_{12}^0$ collisions between the condensate and thermal cloud in
 equilibrium. The solid line plots a direct evaluation of 
(\ref{eq:gam012}), while the 
 circles shows a Monte Carlo calculation for the ``out" rate (solid) and
``in" rate (open).}
 \label{fig:eqm-c12}
\end{figure}

The Monte Carlo calculation of these rates involves a dynamical 
simulation of a sample of test particles moving in the equilibrium 
effective potential. The collisionless evolution of the particles 
in time provides an ergodic sampling of phase space. 
At each time step $\Delta t$ in the evolution, the
collision probabilities in (\ref{eq:22-prob}), (\ref{eq:out-prob}) and
(\ref{eq:in-prob}) are calculated and summed to obtain a realization of
the collision rates at a particular instant of time $t_n$. 
For example, for $C_{22}$ collisions we have
\begin{displaymath}
\Gamma_{22}^0(t_n) \simeq 2 \sum_{(ij)} {P_{ij}^{22}(t_n)
\over \Delta^3 r \Delta t}\,,
\end{displaymath}
where the sum extends over all pairs of test particles in the cell of
volume $\Delta^3 r$.
By repeating this calculation over $M$ time steps and performing the
average
\begin{displaymath}
\langle \Gamma_{22}^0\rangle={1\over M} \sum_{n=1}^M 
\Gamma_{22}^0(t_n)\,,
\end{displaymath}
we obtain the Monte Carlo estimate of the collision rate. The same
procedure is used for the $C_{12}$ ``in" and ``out" rates. To obtain
histograms of the collision rate as a function of the radial position
$r$, we bin the individual collision probabilities according to the
positions of the colliding pair. 
The Monte Carlo results presented in Figs.\ \ref{fig:eqm-c22}
and \ref{fig:eqm-c12} were obtained with only $M= 200$ time steps of
size
$\omega_0 \Delta t = 0.002$, which was already sufficient to give good
statistics. A comparison with the direct numerical
calculations shows very good agreement,
the main error arising from estimating
$f(\bp,\br,t)$ in real-time by binning particles into phase-space cells.
This was confirmed by repeating the simulation but calculating the
collision probabilities using the actual equilibrium Bose distribution
(\ref{eq:bose-dist}) rather than the binned approximation to it. One
can try to improve the binned distribution but there is a trade-off 
between using
smaller phase-space cells which would provide a more accurate
representation of the distribution, and larger cells which contain more
particles and thus improve statistics. Our choice of cell size tries to
optimize these opposing requirements.

The main observation to be made about Fig.~3 is that the ``in"
and ``out'' $C_{12}$ rates are very similar, despite the very different
appearance of the probabilities in (\ref{eq:out-prob}) and 
(\ref{eq:in-prob}). Note in particular that these results confirm 
that the ``in" rate is independent of the arbitrary area ${\cal A}_v$
in (\ref{eq:in-prob}). It is of course important to minimize the 
difference between these two rates since any imbalance implies a net 
transfer of atoms between 
the condensate and thermal cloud which should not occur in equilibrium. 
On the other hand, a calculated imbalance partly reflects the fact that
the equilibrium state we start with is not the ``numerical"
equilibirium state that is consistent with the various numerical
approximations being made. In fact, we find that when a full 
simulation is carried out, the system relaxes to a new, slightly 
different equilibrium. In other words, the system automatically adjusts
to compensate for the numerical approximations. Nevertheless, it is
desirable to avoid an imbalance to whatever extent possible.
Taking the collision rate histograms in Fig.~3 and integrating over 
$r$, we find a
discrepancy between the total ``in" and ``out" rates of about 1\%.
This imbalance can be minimized by judicious choice of
the shape of the phase space elements (see Sec.\ \ref{sec:overv}) and
simulation of a larger sample of test particles, but a residual
imbalance is unavoidable. Since the quantities we are interested in, 
such 
as frequencies and damping rates, are weak functions of the number of
condensate atoms, a small residual drift in the condensate number will
not affect our results significantly.

\subsection{Monopole modes}
\label{sec:monopole}

This section presents the main results of the paper, where we simulate
the monopole ``breathing'' mode in an isotropic trap. These calculations
are not motivated by experiment which have yet to be performed in this
geometry.  Rather, we are mainly interested in comparing our
results to previous theoretical approaches for $C_{12}$ and Landau 
damping which have relied on spherical symmetry. It should be emphasized
that our calculations do not face this restriction, though the simple 
geometry does allow us to more readily observe and quantify effects 
ensuing from $C_{22}$ and $C_{12}$ collisions between atoms. 
In fact, as reported elsewhere \cite{jackson02,jackson01b,jackson02b}, our methods
have already been applied successfully to other experiments in 
anisotropic traps, most notably to the study of scissors modes in which
a full 3D simulation is necessary.

\subsubsection{Static thermal cloud approximation}
\label{sec:static}

As an important test of our treatment of $C_{12}$ collisions, we 
evaluate the damping of the monopole condensate mode within the
so-called static thermal cloud approximation discussed by Williams and
Griffin (WG) \cite{williams01a}. In this approximation, one
considers the dynamics of the condensate in the presence of a static
equilibrium distribution of thermal atoms. Due to the condensate
oscillation, the condensate is no longer in local equilibrium with the
noncondensate and as a result, $C_{12}$ collisions play a role in
damping the mode. This effect enters through the $R$ term in the
generalized GP equation (\ref{eq:GP-gen}). It should be emphasized that
$R$ is provided by the theory and the relaxation it gives rise to is 
not introduced in a phenomenological way as is
sometimes done \cite{pitaevskii59,choi98}. Linearization of the
GP equation leads to generalized Bogoliubov equations which can be 
solved to determine collective mode frequencies and damping rates. The 
latter are of particular interest since they are directly related to the
transfer of atoms between the condensate and thermal cloud as a result 
of
$C_{12}$ collisions. The results obtained \cite{williams01a} are in fact
close to those found in the TF approximation which gives the damping 
rate \cite{williams01c}
\begin{displaymath}
\gamma_j = {\hbar \over 2} {\int d\br \delta n_j^2(\br)/\tau'(\br) \over
\int d\br \delta n_j^2(\br)}\,,
\end{displaymath}
where $\delta n_j(\br)$ is the density fluctuation associated with the
mode $j$ and $1/\tau' = g\Gamma_{12}^0/k_B T$. One sees that the damping
in the TF approximation
is given by a weighted average of the equilibrium $C_{12}$ collision
rate.

Our simulation of the static thermal cloud approximation involves the
propagation of the condensate wavefunction according to 
(\ref{eq:GP-gen}) but with a stationary noncondensate mean field, 
$2g\tilde n_0(r)$. At the same time, the thermal atoms evolve in an 
effective potential defined by the condensate and noncondensate 
equilibrium densities. Although the thermal atoms are not allowed to
undergo collisions, their dynamical evolution allows one to perform 
a Monte Carlo sampling of phase space in order to generate the 
$C_{12}$ collision probabilities at each time step. These probabilites
are then used to calculate the imaginary term, $R(\br,t)$, in the GP 
equation according to (\ref{eq:R-calc}). These simulations can be
compared directly with the calculations by WG 
\cite{williams01a} and therefore provide a direct test of our simulation
methods, in particular, the calculation of $C_{12}$ collision
probabilities. It is important to quantify the errors that arise since
they will also enter into our full simulations in which the effects of
mean fields and collisions on the thermal cloud are included completely.

\begin{figure}
\centering
\psfig{file=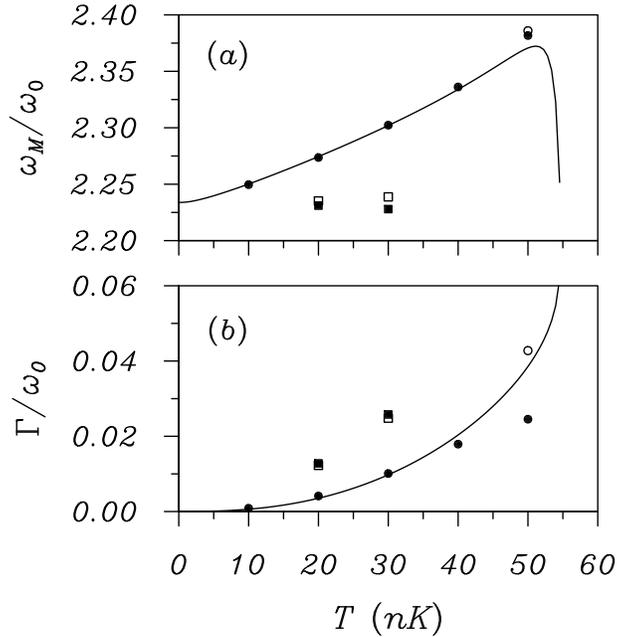, scale=0.43, bbllx=30, bblly = 115,
 bburx=570, bbury=690} 
\caption{Temperature dependent (a) frequency shifts and (b) damping rates of
 the condensate monopole mode in a spherical
 trap ($\omega_0=2\pi \times 10\, {\rm Hz}$), in the presence of a static 
 thermal cloud. The total number of atoms is $N_{\rm tot}=2 \times 10^6$.
 The critical temperature for the corresponding ideal gas is
 $T_0^c = 56.8\, {\rm nK}$. Our results
 are plotted as solid circles, while the solid line is the prediction of
 Williams and Griffin \protect\cite{williams01a}. The open circle at
 $T=50$ nK is the 
 result of a calculation using the analytical form for the
 phase-space density (\ref{eq:bose-dist}). The squares plot results of
 simulations including thermal cloud dynamics, with $C_{22}$ collisions only
 (open) and both $C_{22}$ and $C_{12}$ collisions (closed).} 
\label{fig:static}
\end{figure}

The monopole mode is excited by initially scaling the
equilibrium condensate wavefunction, $\Phi(r,0) = \alpha^{-3/2}
\Phi_0(r/\alpha)$, where the scale parameter $\alpha$ is 0.95.
This dilation of the wavefunction is an alternative to imposing an
initial velocity field as discussed in Sec.~\ref{sec:overv}.
The widths of the condensate wavefunction in the $x$, $y$,
and $z$ directions are defined by mean-squared deviations, e.g.
$\sigma_x \equiv \sqrt{\langle x^2 \rangle - \langle x \rangle^2}$,
where the  moments are given by $\langle \chi \rangle = 
{1\over N_c} \int {\rm d}\br \chi n_c (\br)$. Plots of these
widths show a damped oscillation, and to quantify the frequency $\omega$
and damping rate $\Gamma$, we fit the data to an exponentially decaying 
sinusoid. Since each direction gives slightly different values due to 
statistical fluctuations, we average over the three to obtain values for
$\omega$ and $\Gamma$. Our numerical results are plotted with those of
Ref.\ \cite{williams01a} in Fig.\ \ref{fig:static}. We find excellent 
agreement between the two approaches, except for the damping rate at 
$T = 50\, {\rm nK}$ which is somewhat lower than the WG result. 
This discrepancy arises through errors in estimating the phase-space 
density in the condensate surface region where the fugacity approaches 
unity and the distribution function
$f$ is sharply peaked in momentum space around $\bp=0$. The
$C_{12}$ collision rate in this region is similarly enhanced, especially
at higher temperatures. Our binning
procedure is of insufficient accuracy to fully capture this peak, and 
since the surface region is the major contributor
to the $C_{12}$ damping, this then leads to an underestimate of the 
rate. We illustrate this point in Fig.\ \ref{fig:static} by plotting the
result (open circle) of a simulation at $T = 50$ nK which uses the 
analytical expression for $f_0$ in
(\ref{eq:bose-dist}), as opposed to the binned phase-space density.
We now find 
much better agreement with the WG damping result. The generally good
agreement with WG for the frequency and damping rate
confirms that collision rates can be reliably calculated using our
Monte Carlo sampling methods.

Although the binning procedure introduces some minor errors
into our simulations within the static thermal cloud approximation, 
we expect them to be even less important when the
full dynamics of the thermal cloud is included. Due to mean-field
interactions with the condensate, the thermal cloud will be strongly
perturbed in the surface region and the distribution in phase space will
tend to be ``smeared out", making the binning procedure more reliable.
$C_{22}$ collisions compete against this effect by 
rethermalizing the particles to a Bose distribution; however, this can 
only make a significant difference if the collisional timescale is short
compared to that of the oscillation. 
For the present calculations, we have $\omega_0 \bar{\tau}_{22} >> 1$
and the gas is in the collisionless regime. We would therefore expect
the thermal cloud dynamics to be very important in determining the 
damping due to $C_{12}$ collisions. 

To illustrate this we have 
performed full simulations including mean-field interactions and
collisions at $T= 20$ nK and 30 nK. The results obtained with only 
$C_{22}$ collisions included are shown by open squares, while the
results including $C_{12}$ collisions as well are shown by the full
squares. One sees that the overall damping rate increases by only
5-10\% when $C_{12}$ collisions are added in. In fact, collisions of
either kind contribute little to the damping which is dominated by
Landau damping (as discussed in the following subsection). 
Furthermore, we find a small downward shift in the 
frequency compared to the zero temperature value, in contrast to the
significant increase seen within the static approximation. This increase
is due to the fact that the condensate is oscillating in the presence of
the static mean field of the equilibrium thermal cloud which effectively
enhances the oscillator frequency of the trap. This effect is eliminated
when the thermal cloud is allowed to respond to the dynamic mean field
of the condensate. As we shall
further demonstrate in the following subsection, dynamic mean-field
effects typically dominate the
finite temperature behavior, with collisions playing a secondary but
important supporting role in equilibrating the system. 

As regards the status of the static thermal cloud approximation, we
conclude that it provides a useful method for qualitatively determining
the effects of $C_{12}$ collisions on collective modes. However, its
quantitative predictions for mode frequencies and damping rates are
unreliable.

\subsubsection{Landau damping}
\label{sec:landau}

\begin{figure}
\centering
\psfig{file=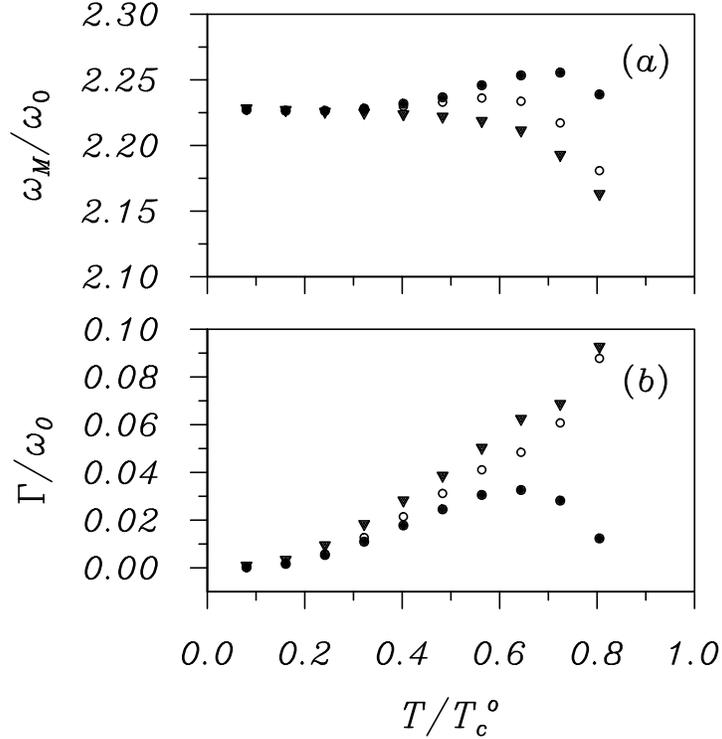, scale=0.5, bbllx = 30, bblly = 115,
 bburx=570, bbury=690}
\caption{(a) Frequency and (b) damping rate of a monopole mode 
 ($\omega_0 = 2\pi \times 187\, {\rm Hz}$, $N_{\rm tot}=5\times 10^4$), 
 including 
 thermal cloud dynamics. Results are shown for simulations with
 no collisions, $C_{22}=C_{12}=0$ (closed circles), $C_{22}$ collisions 
 only (open circles), and both $C_{12}$ and $C_{22}$ collisions (inverted
 triangles).}
\label{fig:monopole} 
\end{figure}

As our final example, we have performed simulations for the system 
studied by Guilleumas and Pitaevskii \cite{guilleumas00}, namely
$^{87}{\rm Rb}$ atoms confined in an isotropic trap of frequency 
$\omega_0=2\pi\times 187\, {\rm Hz}$. 
To begin, we consider a total of $N= 5\times 10^4$ atoms and excite
the monopole mode by an initial scaling of the condensate
radius by a factor of $\alpha = 0.9$, with the thermal cloud 
initially in its equilibrium state.  The condensate 
width oscillations are then followed over a time scale of 
$\omega_0 t =30$. Fig.\ \ref{fig:monopole} shows damping rates and 
frequencies as a function of temperature found by fitting an 
exponentially decaying sinusoid to the time-dependent width. At each 
temperature three simulations are performed. The first
involves free propagation of thermal test particles without collisions, 
corresponding to solving the collisionless Boltzmann equation. The second
includes $C_{22}$ collisions between thermal atoms, while the third 
includes
both $C_{22}$ and $C_{12}$ collisions. At low temperatures, all three
simulations give similar results, reflecting the fact that the number of
thermal atoms is small and collisions play a minor role. With increasing
temperature, the differences between the simulations increase.
Qualitatively, the behaviour is similar to what was found previously
for the scissors mode \cite{jackson01b}; collisions have the effect of shifting the
frequency downward as compared to the collisionless result, and
significantly enhance the damping rate. The effect of $C_{22}$
collisions is particularly strong at high temperatures, which at first
sight may seem surprising since $C_{22}$ collisions do not couple to
the condensate directly.

\begin{figure}
\centering
\psfig{file=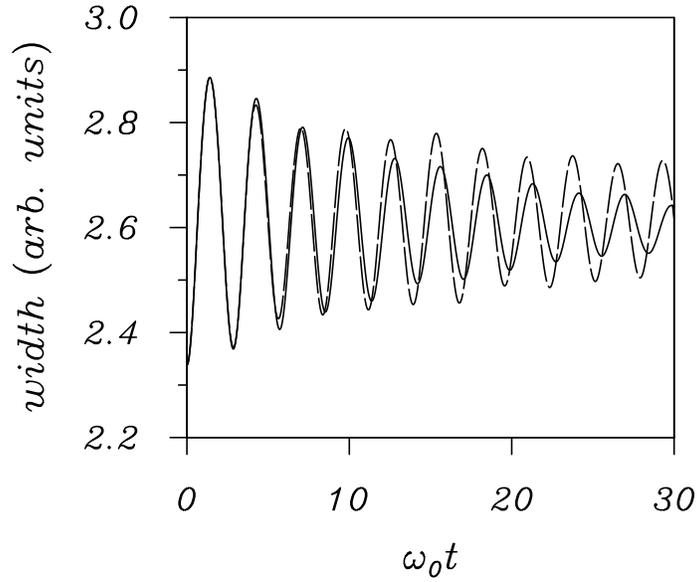, scale=0.5, bbllx = 47, bblly = 120,
 bburx=570, bbury=580}
\caption{Time-dependent width of the condensate, $\sigma_x$, after excitation
 of the monopole mode at $T=200\, {\rm nK}$. The dashed line shows the 
 collisionless evolution, while the result of a full simulation 
 ($C_{12}$ and $C_{22}$) is indicated by the solid line.}
\label{fig:timecurve}
\end{figure}

\begin{figure}
\centering
\psfig{file=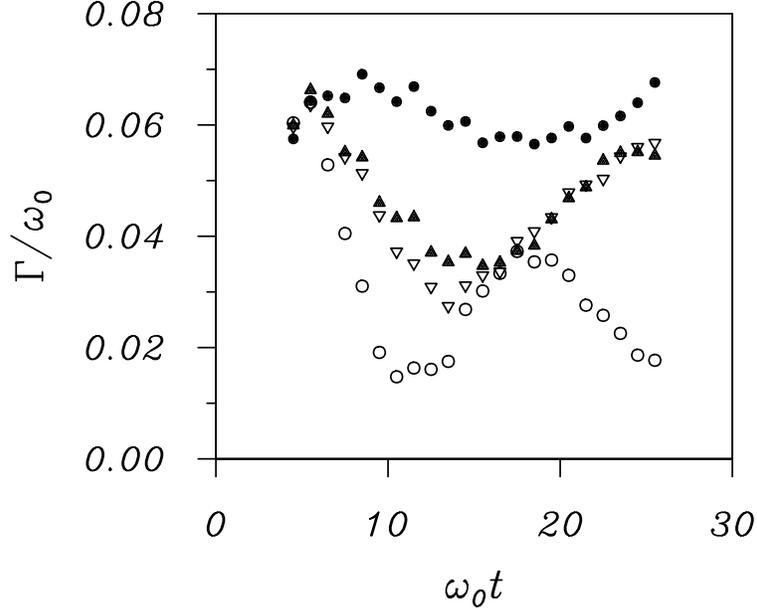, scale=0.53, bbllx = 30, bblly = 120,
 bburx=570, bbury=570}
\caption{Damping rates for the same parameters as in Fig.\ 
 \ref{fig:timecurve}, where fits are taken with a series of windows in the 
 range $[\omega_0 t-4.5,\omega_0 t+4.5]$. We plot data for 
 simulations which include no collisions (open circles), $C_{12}$ only 
 (inverted triangles), $C_{22}$ only (solid triangles), and both $C_{12}$ 
 and $C_{22}$ collisions (solid circles).}
\label{fig:dampcurve}
\end{figure}

To gain more insight into the collisional dependence of the damping, 
we focus on the time-dependent evolution for a particular 
temperature, $T=200\, {\rm nK}$ 
($T/T_c^0 = 0.644$, where $T_c^0=0.94\hbar\omega_0 N_{\rm tot}^{1/3}$ 
\cite{dalfovo99} is the transition temperature of the corresponding ideal gas
in the thermodynamic limit). Fig.\ \ref{fig:timecurve} plots $\sigma_x$ {\it 
vs.} $t$ for the collisionless and full ($C_{12}$+$C_{22}$) simulations.
The initial
damping rate in both calculations is seen to be similar, however at later
times the collisionless oscillation departs from a simple exponential 
decay, and the oscillation amplitude tends to saturate. This behavior is not seen to 
the same degree when collisions are included. To quantify this behavior,
we define a local damping rate by fitting a damped sinusoid to the data
within a window of width $\Delta(\omega_0 t) = 9$ centered on the time
$t$.  Fig.\ \ref{fig:dampcurve} plots this local damping rate as a
function of $t$.
We see large variations in the damping rate with time, with the largest
rate occuring initially. The deviations from the initial value are
largest in the collisionless case, where the damping rate
dips nearly to zero.   
Similar behavior is observed over the whole range of temperatures, and 
accounts for the lower damping rates obtained by fitting the 
entire data set.   

To explain this behavior, we note that damping of the condensate
oscillation is associated with
the transfer of energy from the condensate to the 
thermal cloud. If this energy exchange is mediated by mean-field 
interactions, it is referred to as 
Landau damping. From the point of view of the thermal cloud, the
dynamic condensate mean field $2g n_c(\br,t)$ acts as an external
perturbation which can lead to the excitation of thermal atoms. 
Of course, the rate at which these excitations occur
depends on the phase-space distribution of the thermal particles.
In our simulations, the thermal cloud is initially in an
equilibrium state and the damping rate is observed to be independent of
collisions. This damping is essentially pure Landau damping and its
magnitude is determined by the rate at which the oscillating condensate
can do work on the equilibrium thermal distribution. In this respect,
our initial damping rate is analogous to conventional perturbation 
theory estimates (as discussed below). 
As time progresses in our simulations, the thermal
cloud begins to deviate from an equilibrium distribution and the
magnitude of Landau damping is correspondingly affected. Evidently, the
perturbation of the thermal distribution is such as to reduce the rate
of energy transfer to the thermal atoms, whereupon the damping rate 
decreases with time as seen in Fig.\ \ref{fig:dampcurve}.
The deviation is in fact a nonlinear effect as it was found to depend on
the amplitude of the condensate oscillation.
An analogous effect appears in the context of plasma 
oscillations, where Landau damping is due to the energy transfer from
the collective plasma wave to single-electron excitations 
\cite{stix62,platzman61}.

In the absence of collisions, the
distribution of thermal atoms continues to evolve in a complicated way
and the effective damping rate exhibits an oscillatory time dependence.
However, as soon as $C_{22}$ collisions are switched on, the damping
rate deviates less strongly from its initial value. The effect of these
collisions is to drive the thermal cloud towards a state of local
equilibrium and the damping rate tends to maintain its original
value. The inclusion of $C_{12}$ collisions has a similar effect and we
find a damping rate which is almost time-independent when both collision
processes are retained. However, $C_{12}$ collisions do more than simply
equilibrate the thermal cloud since they also lead to the source term
$R(\br,t)$ in the GP equation. As we have already discussed, this term
gives rise to its own contribution to damping which is quite
separate from Landau damping. It should be emphasized  that it
is impossible to separate the total damping rate into individual
components. Mean-field and collisional effects are interrelated,
and all must be included to completely account for the actual damping
rates.

We next turn to a comparison of our results with those of Guilleumas and
Pitaevskii \cite{guilleumas00}. Since we have used quite different
methods to calculate damping rates, it is useful to first discuss the
perturbation theory calculation of Landau damping used by these 
authors \cite{pitaevskii97}. Within this approach, Landau damping is
associated with the decay of a mode of oscillation (with energy $\hbar
\omega_{\rm osc}$) as a result of the excitation of a thermal
quaisparticle from an initial state of energy $E_i$ to a final state of
energy $E_k$. The damping rate is
then given by Fermi's golden rule 
\cite{guilleumas00,pitaevskii97,giorgini98,giorgini00}
\begin{equation}
 \Gamma = \frac{\pi}{\hbar} \sum_{ik} |A_{ik}|^2 [f (E_i) - f (E_k)]
 \delta (E_k - E_i - \hbar \omega_{\rm osc}),
\label{eq:perturb}
\end{equation}
where the sum is over all excitations that satisfy energy conservation, 
while the matrix element $A_{ik}$ depends upon the form of the 
excitations.  The thermal states are occupied according to the
equilibrium Bose distribution $f(E)$. This damping rate is therefore
analogous to the {\it initial} damping rate we obtain in simulations 
which start with an equilibrium thermal distribution. 

Guilleumas and Pitaevskii \cite{guilleumas00} evaluate 
(\ref{eq:perturb}) as a function of temperature using Bogoliubov 
excitations of the condensate as the thermal quasiparticles. These are
determined from the Bogoliubov equations 
for a fixed number of condensate atoms, $N_c$; the corresponding 
number of thermal atoms is then a function of temperature and is given 
by summing over the thermal occupation $f(E_i)$ of the quasiparticle
states. To actually evaluate the
Landau damping rate at the frequency of the monopole mode of interest,
the delta functions in (\ref{eq:perturb}) are replaced by Lorentzians of
width $\Delta$. They show that the results obtained are essentially
independent of this parameter.

\begin{figure}
\centering
\psfig{file=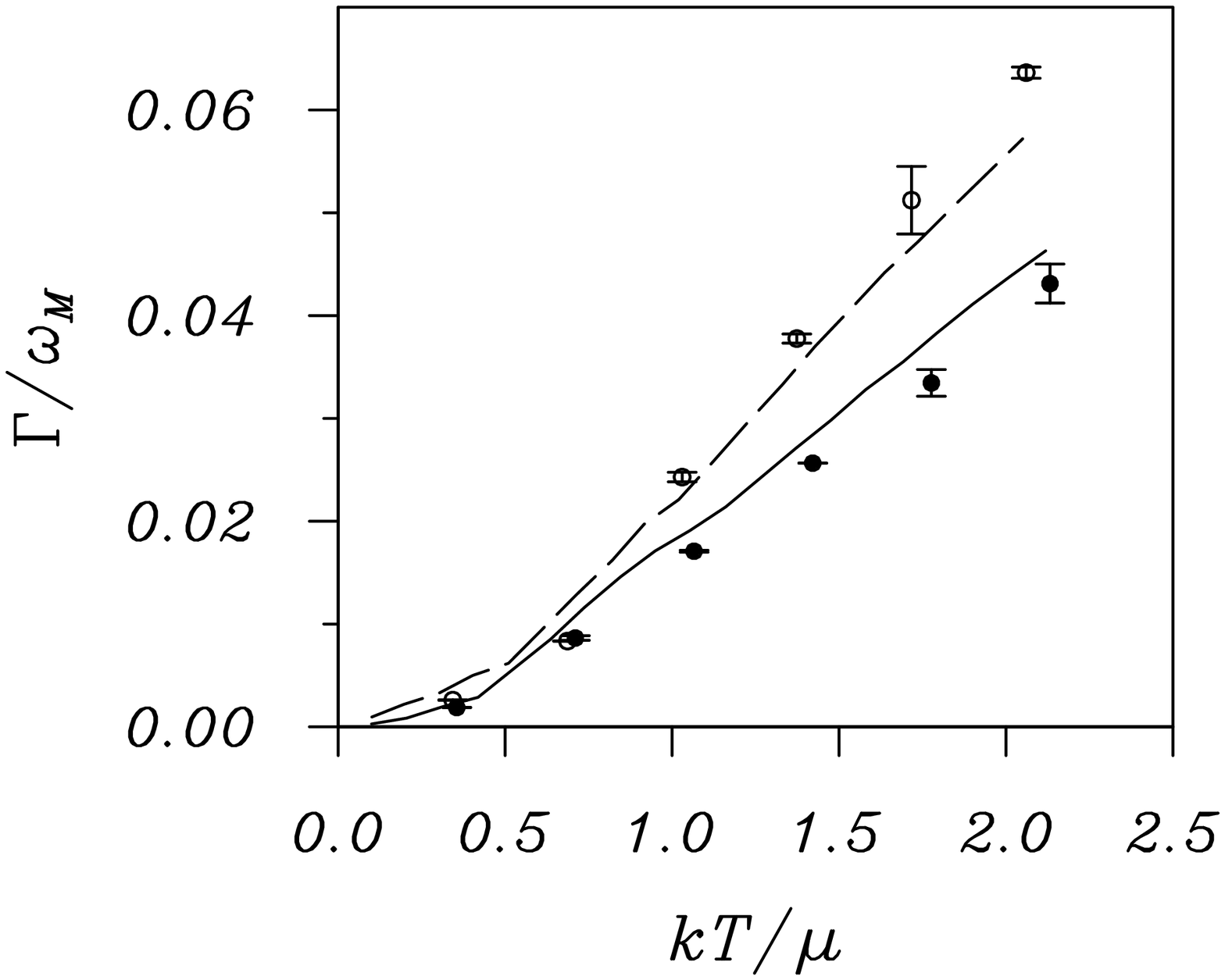, scale=0.53, bbllx = 30, bblly = 120,
 bburx=575, bbury=560}
 \caption{Initial damping rate calculated over the first time interval
 of size $\Delta(\omega_0 t)=9$ (points) compared
 to the results of Guilleumas and Pitaevskii \protect\cite{guilleumas00} 
 (lines). Results are plotted for $N_c = 5 \times 10^4$
 (solid points and line) and $N_c = 1.5 \times 10^5$ (open points, dashed
 line) condensate atoms. 
 Following Ref.\ \protect\cite{guilleumas00} quantities are plotted in terms 
 of dimensionless units, with the ratio $\Gamma/\omega_M$ (damping rate over
 mode frequency) plotted against $k_B T / \mu$. For $\Gamma/\omega_M$ we 
 calculate the mean over the three directions, while the standard deviation
 yields a rough estimate of the error.}   
\label{fig:gupi}
\end{figure}

To compare with these results we performed collisionless simulations and
extracted the initial damping rate as discussed earlier. The comparison
is made in Fig.~\ref{fig:gupi} where results are presented as a function of
temperature for $N_c = 5\times 10^4$ and $1.5 \times 10^5$. Given the
completely different methods of calculation, the agreement is
remarkable. The agreement persists even down
to low temperatures where one might expect differences to appear as a
result of our use of semiclassical HF excitations as opposed 
to the Bogoliubov excitation spectrum. The fact that the agreement is as
good as it is is
perhaps understandable in view of the observation in Ref.\
\cite{you97,dalfovo97} that the density of states in the HF and
Bogoliubov approximations are very similar. Although the semiclassical
HF approximation was not discussed, we would of course expect it to be
close to the quantal HF result. Since the density of thermal excitations is an
important ingredient in the calculation of Landau damping, we can begin
to see why our semiclassical calculations give very similar results.

\section{Conclusions}

In this paper we have provided a detailed description of the numerical 
scheme we have used to simulate trapped 
Bose-condensed gases at finite temperatures, based on the ZGN formalism
which treats the thermal excitations semiclassically within a 
Hartree-Fock-Popov approximation. The procedure involves solving 
simultaneously a generalized Gross-Pitaevskii 
equation for the condensate and a
Boltzmann kinetic equation for the thermal cloud. The two equations are
coupled by mean fields and collisions, both of which influence the 
dynamics of the two components in significant ways. Our scheme has been
carefully tested to ensure that it provides an accurate description of
the system dynamics. In particular, we have shown that $N$-body 
simulations, together with the Monte Carlo sampling of collisions, 
is an effective and reliable method for determining the thermal cloud 
dynamics. 

Our scheme can be 
used to model the dynamics of the gas over a wide range of temperatures
and physical conditions. As an example, we have studied the monopole 
``breathing'' mode in a spherical trap. Two sets of calculations were
performed. The first provided a check of the treatment of collisions
within the static thermal cloud approximation of Williams and Griffin
\cite{williams01a}.
Results within this approximation were reproduced, but full dynamical
simulations indicated that the approximation is primarily useful as a
qualitative indicator of the effect of $C_{12}$ collisions.
Unfortunately, its quantitative predictions for mode frequencies and
damping rates cannot be trusted. Our second set of simulations focussed
on Landau damping. This is typically the dominant damping process for
condensate modes at finite temperatures. However, the damping observed
in a simulation, and by extension in real experimental situations, is
determined by a delicate interplay of the mean-field excitation of the
thermal cloud and collisions. The thermalizing effect of the latter
strongly influences the rate at which mean-field excitations take place.

We also compared our results for Landau damping 
to those of Guilleumas and Pitaevskii \cite{guilleumas00}, and 
very good agreement was found. This confirms that the semiclassical
HF description of the thermal cloud reproduces the Landau
damping as calculated using Bogoliubov excitations. This is not too
surprising since the density of excitations in the two approximations is
very similar. However, as we have already explained, the Landau damping
as determined by assuming the thermal cloud to be in an equilibrium 
state is not necessarily the damping that will be observed in an
experiment. A consistent treatment of the dynamics of the
condensate {\it and} thermal cloud is needed in order to make detailed
comparisons with experiment.

Applications of our technique to other 
collective modes have also been made and are discussed elsewhere 
\cite{jackson02,jackson01b}. Our results for the temperature-dependent
damping and frequency shifts are in 
good agreement with experiment for both quadrupole
\cite{jackson02} and scissors \cite{jackson01b} modes. This in itself
confirms the accuracy of our theoretical formulation of the system
dynamics and the numerical methods used. It is hoped, however, that the
present paper provides more insight into the content of the theory and
the reasons for its success. Interesting future systems for
study could include topological defects (e.g.\ vortices and skyrmions),
optical lattices, and dynamical instabilities of surface modes in the 
presence of a rotating thermal cloud \cite{williams01b}. 

We conclude with a few comments about where we may go next. It would be
useful to incorporate Bogoliubov excitations in the kinetic theory in
place of HF excitations \cite{imamovic01}. 
Although this does not seem to be important for
Landau damping, the HF approximation may not be good for certain
situations where the thermal occupation of the lowest modes are
significant (e.g., for large atom numbers or at low temperatures). 
Another possibility is 
to implement a hybrid scheme, where highly-occupied,
low-lying modes are treated using
classical field methods \cite{sinatra01,davis01}, 
while the rest are treated 
semiclassically using the present technique. Finally, it would be of
interest to investigate the importance of going
beyond the Popov approximation by including the anomalous average,
$\tilde{m}$. In doing so one must be careful to ensure that the new 
model is gapless \cite{griffin96}.
This may involve renormalization of the coupling
constant $g$ \cite{hutchinson98} or replacing the contact potential by a
generalized pseudopotential \cite{olshanii01}.      

\begin{acknowledgements}
We thank J. Williams for providing the data in Fig.\ \ref{fig:static}. 
We would also like to acknowledge useful discussions with A. Griffin, T.
Nikuni and J. Williams.
Financial support was provided by NSERC of Canada.
\end{acknowledgements}

\appendix
\section{Monte Carlo Calculation of Collision Rates}

Our purpose here is to show how a Monte Carlo evaluation of the
collision rate in (\ref{gamma22_out}) leads to a definition of 
collision probabilities
to be used in the simulations. We first note that $d^3r d^3p/h^3
C_{22}^{\rm out}$ represents the number of atoms leaving the phase 
space volume element $d^3r d^3p/h^3$ per unit time as a result of 
collisions.
Integrating this over momenta gives the number of atoms in $d^3r$
suffering a collision per unit time. Thus the mean collision rate per
atom and per unit volume is
\begin{eqnarray}
\Gamma_{22}^{\rm out} = \int {d^3p \over h^3}  C_{22}^{\rm out}\,,
\end{eqnarray}
which is the quantity displayed in (\ref{eq:22-rate}).
We now write the required local collision rate as
\begin{eqnarray}
\Gamma_{22}^{\rm out} &=& \int {d^3p_1 \over h^3}  \int {d^3p_2 \over h^3}
f(\bp_1) f(\bp_2) g(\bp_1, \bp_2)\nonumber \\
&\equiv& \int d^6 p\, w(p) g(p)
\end{eqnarray}
where $p$ is a point in 6-dimensional momentum space and the factor 
$w(p) \equiv f(\bp_1)f(\bp_2)/h^6$ is considered as a weight function. 
We denote the maximum value of $w(p)$ by $w_{\rm max}$ and 
define the domain on which the integrand is nonzero by
$[-p_{\rm max}/2,\,p_{\rm max}/2]$ for
each momentum component. Choosing a point $p_i$ at random in the
hypervolume $(p_{\rm max})^6$, and a random number $R_i$ uniformly
distributed on $[0, w_{\rm max}]$, the point $p_i$ is accepted if $R_i <
w(p_i)$ and the quantity $g(p_i)$ is accumulated. The value of the
integral is then given approximately as
\begin{equation}
\Gamma_{22}^{\rm out} \simeq (p_{\rm max})^6 w_{\rm max} {1\over N} 
{\sum_i}' g(p_i)\,,
\end{equation}
where $N$ is the number of random $p_i$ points chosen and the prime on
the summation includes only those points for which $R_i < w(p_i)$ .
For $g\equiv 1$, the integral is simply $\tilde n(\br)^2$. Thus,
\begin{equation}
\tilde n(\br)^2 = (p_{\rm max})^6 w_{\rm max} {N_s \over N}
\end{equation}
where $N_s$ is the total number of points accepted, and
\begin{equation}
\Gamma_{22}^{\rm out} \simeq \tilde n^2 {1\over N_s} {\sum_i}' g(p_i)\,.
\end{equation}

The sample of $N_s$ points accepted consists of $N_s$ $\bp_1$-values and
$N_s$ $\bp_2$-values, each of which is distributed according to
$f(\bp)$. This set of $2N_s$ $\bp$-values can be identified with
$N_{\rm cell}$ test particles in a cell of volume $\Delta^3r$.
If this set is to be representative of the local density, we must have
\begin{equation}
\tilde n(\br) = {N_{\rm cell} \over \Delta^3r} = {2N_s \over
\Delta^3r}\,.
\end{equation}
With this identification,
\begin{equation}
\Delta^3 r \Gamma_{22}^{\rm out} = 2 \tilde n \sum_{i=1}^{N_s}
g(\bp_1^i,\bp_2^i)\,.
\end{equation}
In other words, the collision rate can be estimated by sampling the test
particles in the cell $\Delta^3r$ {\it in pairs}. Inserting the explicit
form of $g$ for the 22 collision rate in (\ref{gamma22_out}), we have
\begin{equation}
\Delta^3 r \Gamma_{22}^{\rm out} = 2 \sum_{(ij)}
\tilde n(\br) \sigma |\bv_i - \bv_j| \int {{\rm d}\Omega \over 4\pi}
(1+f_3)(1+f_4)\,,
\label{A8}
\end{equation}
where the sum is now taken over pairs of test particles. This expression
allows us to define the probability $P_{ij}^{22}$ that a pair of
atoms $(ij)$ in the cell suffers a collision in a time interval $\Delta
t$:
\begin{equation}
P_{ij}^{22} =
\tilde n(\br) \sigma |\bv_i - \bv_j| \int {{\rm d}\Omega \over 4\pi}
(1+f_3)(1+f_4)\Delta t\,.
\end{equation}
Selecting atoms in pairs from each cell and assigning them a
collision probability $P_{ij}^{22}$ allows us to simulate the effect of
collisions in a way which is consistent with the Boltzmann collision
integral. Note that the factor of 2 in (\ref{A8}) accounts for the 
fact that 2 atoms are affected for each pair collision. This factor is
therefore not included in the definition of the pair collision
probability.

We treat $C_{12}$ collisions somewhat differently. First, we note that
the total rate of change of the number of thermal atoms per unit volume 
due to these collisions is
\begin{eqnarray}
\int {d^3 p \over h^3} C_{12} &=& 
\frac{\sigma n_c}{\pi m^2 h^3} \int {\rm d}\bp_2 {\rm d}\bp_3
 {\rm d}\bp_4 \delta(m\bv_c + \bp_2-\bp_3-\bp_4)
 \delta(\epsilon_c+\epsilon_2-\epsilon_3-\epsilon_4) \nonumber \\
 &&\qquad \times [f_2 (1+f_3)(1+f_4) - (1+f_2)f_3 f_4 ]\nonumber \\
&\equiv& \Gamma_{12}^{\rm out} - \Gamma_{12}^{\rm in}\,.
\end{eqnarray}
According to this definition,
\begin{eqnarray}
\Gamma_{12}^{\rm out} &=& 
\frac{\sigma n_c}{\pi m^2 h^3} \int {\rm d}\bp_2 {\rm d}\bp_3
 {\rm d}\bp_4 \delta(m\bv_c + \bp_2-\bp_3-\bp_4)
 \delta(\epsilon_c+\epsilon_2-\epsilon_3-\epsilon_4) f_2(1+f_3)(1+f_4)
\nonumber \\
&=& \int {d^3 p_2 \over h^3} f_2 n_c \sigma v_r^{\rm out} \int
{d\Omega \over 4\pi} (1+f_3)(1+f_4)
\label{Agamout}
\end{eqnarray}
is the rate of decrease of the number of {\it condensate} atoms per unit
volume as a result of a collision with a thermal atom (hence the
designation `out'). This rate can be
estimated by writing
\begin{equation}
\Gamma_{12}^{\rm out} = 
\int {\rm d}^3 p_2 w(p) g(p)\,,
\label{A12}
\end{equation}
where $w(p) = f(p)/h^3$ and $g(p)$ is the remaining part of the
integrand. A Monte Carlo sampling of the integral leads to the
estimate
\begin{equation}
\Delta^3 r \Gamma_{12}^{\rm out} \simeq \sum_{i=1}^{N_s} g(p_i)\,,
\label{MC-out}
\end{equation}
where $N_s$ represents the number of atoms in the cell of 
volume $\Delta^3 r$.
The probability of an atom in the cell suffering this kind of collision
in the time interval $\Delta t$ is therefore
\begin{equation}
P_i^{\rm out} = g(p_i) \Delta t = n_c \sigma 
\sqrt{|\bv_c - \bv_2^i|^2 - 4gn_c/m}\int {{\rm d}\Omega
\over 4\pi} (1+f_3)(1+f_4)\Delta t\,,
\end{equation}
which is the origin of the expression given in (\ref{eq:out-prob}).

The `in' collision rate is given by
\begin{eqnarray}
\Gamma_{12}^{\rm in} &=& 
\frac{\sigma n_c}{\pi m^2 h^3} \int {\rm d}\bp_2 {\rm d}\bp_3
 {\rm d}\bp_4 \delta(m\bv_c + \bp_2-\bp_3-\bp_4)
 \delta(\epsilon_c+\epsilon_2-\epsilon_3-\epsilon_4) (1+f_2) f_3 f_4
\nonumber \\
&=& \int {d^3 p_2 \over h^3} f_2 \int {d^3 p_4 \over h^3} f_4 
{n_c \sigma h^3 \over \pi m} \delta [ (\bp_c - \bp_4)\cdot (\bp_c -
\bp_2) - mgn_c] (1+f_3)\,,
\label{Agamin}
\end{eqnarray}
where we have interchanged the particle labels 2 and 3 to obtain the
second line in this equation.
This rate corresponds to two thermal atoms
scattering into a condensate atom and an outgoing thermal atom, and is
thus the rate that atoms feed into the condensate as a result of
collisions. Although the collision of atoms 2 and 4 can be treated by
the methods used to analyze the $C_{22}$ collision rate, it is
preferable to define a single atom collision rate by writing this
integral in the form of (\ref{A12}) and performing a Monte Carlo
sampling with respect to the $\bp_2$ variable. This procedure 
leads to the  collision probability {\it per atom}
\begin{equation}
P_i^{\rm in} = \Delta t \int {d^3 p_4 \over h^3} (1+f_3) f_4 
{n_c \sigma h^3 \over \pi m} \delta [ (\bp_c - \bp_4)\cdot (\bp_c -
\bp_2^i) - mgn_c]\,,
\end{equation}
which is simplified and discussed further in the body of the paper.

\begin{references}

\bibitem{anderson95} M. H. Anderson, J. R. Ensher, M. R. Matthews, C. E.
 Wieman, and E. A. Cornell, Science {\bf 269}, 198 (1995). 
\bibitem{bradley95} C. C. Bradley, C. A. Sackett, J. J. Tollett, and R.
 G. Hulet, Phys. Rev. Lett. {\bf 75}, 1687 (1995).
\bibitem{davis95} K. B. Davis, M. O. Mewes, M. R. Andrews, N. J. van
 Druten, D. S. Durfee, D. M. Kurn, and W. Ketterle, Phys. Rev. Lett. 
 {\bf 75}, 3969 (1995).
\bibitem{dalfovo99} F. Dalfovo, S. Giorgini, L. P. Pitaevskii, and 
 S. Stringari, Rev. Mod. Phys. {\bf 71}, 463 (1999).
\bibitem{hutchinson97} D. A. W. Hutchinson, E. Zaremba, and A. Griffin,
 Phys. Rev. Lett. {\bf 78}, 1842 (1997).
\bibitem{dodd98} R. J. Dodd, M. Edwards, C. W. Clark, and K. Burnett,
 Phys. Rev. A {\bf 57}, R32 (1998).
\bibitem{griffin96} A. Griffin, Phys. Rev. B {\bf 53}, 9341 (1996).
\bibitem{jin97} D. S. Jin, M. R. Matthews, J. R. Ensher, C. E. Wieman, 
 and E. A. Cornell, Phys. Rev. Lett. {\bf 78}, 764 (1997).
\bibitem{stamper-kurn98} D. M. Stamper-Kurn, H. -J. Miesner, S. Inouye, 
 M. R. Andrews, and W. Ketterle, Phys. Rev. Lett. {\bf 81}, 500 (1998). 
\bibitem{marago01} O. M. Marag\`{o}, G. Hechenblaikner, E. Hodby, and 
 C. J. Foot, Phys. Rev. Lett. {\bf 86}, 3938 (2001).
\bibitem{chevy01} F. Chevy, V. Bretin, P. Rosenbusch, K. W. Madison, and
 J. Dalibard, cond-mat/0111455.
\bibitem{morgan00} S. A. Morgan, J. Phys. B {\bf 33}, 3847 (2000).
\bibitem{rusch00} M. Rusch, S. A. Morgan, D. A. W. Hutchinson, and K. 
 Burnett, Phys. Rev. Lett. {\bf 85}, 4844 (2000).
\bibitem{giorgini98} S. Giorgini, Phys. Rev. A {\bf 57}, 2949 (1998).
\bibitem{giorgini00} S. Giorgini, Phys. Rev. A {\bf 61}, 063615 (2000).
\bibitem{reidl00} J. Reidl, A. Csord\'{a}s, R. Graham, and P. 
Sz\'{e}pfalusy, Phys. Rev. A {\bf 61}, 043606 (2000).
\bibitem{gardiner00} C. W. Gardiner and P. Zoller, Phys. Rev. A {\bf 61},
 033601 (2000), and references therein.
\bibitem{stoof99} H. T. C. Stoof, J. Low Temp. Phys. {\bf 114}, 11
(1999).
\bibitem{walser99} R. Walser, J. Williams, J. Cooper, and M. Holland, Phys.
 Rev. A {\bf 59}, 3878 (1999).
\bibitem{nikuni99} T. Nikuni, E. Zaremba, and A. Griffin, Phys. Rev.
Lett. {\bf 83}, 10 (1999).
\bibitem{zaremba99} E. Zaremba, T. Nikuni, and A. Griffin, J. Low Temp.
 Phys. {\bf 116}, 277 (1999).
\bibitem{nikuni01a} T. Nikuni and A. Griffin, Phys. Rev. A {\bf 63},
 033608 (2001).
\bibitem{bijlsma99} M. J. Bijlsma and H. T. C. Stoof, Phys. Rev. A {\bf
 60}, 3973 (1999).
\bibitem{alkhawaja00} U. Al Khawaja and H. T. C. Stoof, Phys. Rev. A 
 {\bf 62}, 053602 (2000).
\bibitem{nikuni01b} T. Nikuni, Phys. Rev. A {\bf 65}, 033611 (2002).
\bibitem{guery-odelin99} D. Guery-Odelin and S. Stringari, Phys. Rev.
 Lett. {\bf 83}, 4452 (1999).
\bibitem{marago00} O. M. Marag\`{o}, S. A. Hopkins, J. Arlt, E. Hodby, 
 G. Hechenblaikner, and C. J. Foot, Phys. Rev. Lett. {\bf 84}, 2056 
 (2000).
\bibitem{jackson02} B. Jackson and E. Zaremba, Phys. Rev. Lett. {\bf
 88}, 180402 (2002).
\bibitem{jackson01b} B. Jackson and E. Zaremba, Phys. Rev. Lett. {\bf 87}, 
 100404 (2001).
\bibitem{jackson02b} B. Jackson and E. Zaremba, Laser Phys. {\bf 12}, 93
(2002).
\bibitem{kadanoff89} L. P. Kadanoff and G. Baym, {\it Quantum Statistical
 Mechanics} (Addison-Wesley, Redwood City, 1989).
\bibitem{taha84} T. R. Taha and M. J. Ablowitz, J. Comput. Phys. {\bf 55},
 203 (1984).
\bibitem{sanz-serna94} J. M. Sanz-Serna and M. P. Calvo, {\it Numerical
 Hamiltonian Problems} (Chapman \& Hall, London, 1994).
\bibitem{nr} W. H. Press, S. A. Teukolsky, W. T. Vetterling, and B. P.
 Flannery, {\it Numerical Recipes in FORTRAN} (Cambridge University
 Press, Cambridge, 1992).
\bibitem{footnote1} See e.g.\ \texttt{http://www.fftw.org}.
\bibitem{hockney81} R. W. Hockney and J. W. Eastwood, {\it Computer
 simulations using particles} (McGraw-Hill, New York, 1981).
\bibitem{yoshida93} H. Yoshida, Celest. Mech. Dyn. Astron. {\bf 56}, 27 
 (1993).
\bibitem{prigogine} I. Prigogine, {\it Non-equilibrium Statistical
Mechanics} (Wiley, New York, 1962).
\bibitem{wu97} H. Wu, E. Arimondo, and C. J. Foot, Phys. Rev. A 
 {\bf 56}, 560 (1997).
\bibitem{williams01a} J. E. Williams and A. Griffin, Phys. Rev. A
 {\bf 64}, 013606 (2001).
\bibitem{pitaevskii59} L. P. Pitaevskii, Sov. Phys. JETP {\bf 35} 282
(1959).
\bibitem{choi98} S. Choi, S. A. Morgan and K. Burnett, Phys. Rev. A {\bf
57}, 4057 (1998).
\bibitem{williams01c} J. E. Williams and A. Griffin, Phys. Rev. A
 {\bf 63}, 023612 (2001).
\bibitem{guilleumas00} M. Guilleumas and L. P. Pitaevskii, Phys. Rev. A
 {\bf 61}, 013602 (2000).
\bibitem{stix62} T. H. Stix, {\it The Theory of Plasma Waves} (McGraw-Hill,
 New York, 1962).
\bibitem{platzman61} P .M. Platzman and S. J. Buchsbaum, Phys. Fluids
 {\bf 4}, 1288 (1961).
\bibitem{pitaevskii97} L. P. Pitaevskii and S. Stringari, Phys. Lett. A
 {\bf 235}, 398 (1997).
\bibitem{you97} L. You, W. Hoston, and M. Lewenstein, Phys. Rev. A {\bf 55},
 R1581 (1997).
\bibitem{dalfovo97} F. Dalfovo, S. Giorgini, M. Guilleumas, L. Pitaevskii,
 and S. Stringari, Phys. Rev. A {\bf 56}, 3840 (1997).
\bibitem{williams01b} J. E. Williams, E. Zaremba, B. Jackson, T. Nikuni,
 and A. Griffin, Phys. Rev. Lett. {\bf 88}, 070401 (2002).
\bibitem{imamovic01} M. Imamovi\'c-Tomasovi\'c and A. Griffin, J. Low Temp.
Phys. {\bf 122}, 617 (2001).
\bibitem{sinatra01} A. Sinatra, C. Lobo, and Y. Castin, Phys. Rev. Lett. 
 {\bf 87}, 210404 (2001); cond-mat/0201217.
\bibitem{davis01} M. J. Davis, S. A. Morgan, and K. Burnett, Phys. Rev. Lett.
 {\bf 87}, 160402 (2001); cond-mat/0201571.
\bibitem{hutchinson98} D. A. W. Hutchinson, R. J. Dodd, and K. Burnett, 
 Phys. Rev. Lett. {\bf 81}, 2198 (1998).
\bibitem{olshanii01} M. Olshanii and L. Pricoupenko, Phys. Rev. Lett. 
 {\bf 88}, 010402 (2002). 

\end{references}
\end{document}